\documentclass[iop]{emulateapj}
\usepackage{color}
\usepackage{times}
\usepackage{graphicx}
\usepackage{subfigure}
\usepackage{float}
\usepackage[colorlinks=true,linkcolor=blue,citecolor=blue]{hyperref}
\usepackage{txfonts}
\usepackage{epsfig}
\usepackage{lscape}

\def\co{$^{12}$CO}
\def\tco{$^{13}$CO}
\def\deg{^{\circ}}
\renewcommand{\arraystretch}{2.0}

\newcommand{\hii}{H \textsc{ii}}
\newcommand{\msun}{$M_\odot$}
\journalinfo{Accepted to the Astrophysical Journal}
\submitted{Received 2013 November 1; accepted 2014 October 11}
\shorttitle{Star Formation in the Bubble N6}
\shortauthors{Yuan et al.}
\begin{document}

\title{Expanding Shell and Star Formation in the Infrared Dust Bubble N6}
\author{Jing-Hua Yuan\altaffilmark{1}\footnotemark[\dag], Yuefang Wu \altaffilmark{2}\footnotemark[\ddag], Jin Zeng Li\altaffilmark{1}, and Hongli Liu \altaffilmark{1}}
\affil{$^1$National Astronomical Observatories, Chinese Academy of Sciences, 20A Datun Road, Chaoyang District, Beijing 100012, China;\\}
\affil{$^2$Department of Astronomy, Peking University, 100871 Beijing, China;}
\footnotetext[\dag]{jhyuan@nao.cas.cn}
\footnotetext[\ddag]{ywu@pku.edu.cn}

\begin{abstract}

    We have carried out a multi-wavelength study of the infrared dust bubble N6 to extensively investigate the molecular environs and star-forming activities therein. Mapping observations in \co\ $J=1-0$ and \tco\ $J=1-0$ performed with the Purple Mountain Observatory 13.7-m telescope have revealed four velocity components. Comparison between distributions of each component and the infrared emission suggests that three components are correlated with N6. There are ten molecular clumps detected. Among them, five have reliable detection in both \co\ and \tco\ and have similar LTE and non-LTE masses ranging from 200 to higher than 5,000 $M_\sun$. With larger gas masses than virial masses, these five clumps are gravitationally unstable and have potential to collapse to form new stars. The other five clumps are only reliably detected in \co\ and have relatively small masses. Five clumps are located on the border of the ring structure and four of them are elongated along the shell. This is well in agreement with the collect and collapse scenario. The detected velocity gradient reveals that the ring structure is still under expansion due to stellar winds from the exciting star(s). Furthermore, 99 young stellar objects have been identified based on their infrared colors. A group of YSOs reside inside the ring, indicating active star formation in N6. Although no confirmative features of triggered star formation detected, the bubble and the enclosed \hii\ region have profoundly reconstructed the natal could and altered the dynamics therein.

\end{abstract}
\keywords{H \textsc{ii} regions -- ISM: bubbles -- ISM: clouds -- stars: early-type -- stars: formation -- radio lines: ISM}

\section{Introduction}
    In spite of the relatively small number, massive stars are dominant contributors to ionizing photons and heavy elements in galaxies. Formed mainly in clusters \citep{lad03}, massive stars play a crucial role in the evolution of molecular clouds and star formation therein. In the course of formation and evolution of massive stars, ultraviolet (UV) radiations, intense stellar winds, and supernovae can reconstruct the surrounding interstellar medium to trigger or hinder the formation of new generation of stars.

    In recent years, influences of hot stars on their natal clouds have been more and more tested in bubbles. The successful {\it Spitzer} observations have revealed the bubbling Galactic disk of the Milky Way. Based on the GLIPMSE \citep{ben03,chu09} and MIPSGAL \citep{car09} survey data, the first large catalog of about 600 infrared dust bubbles were identified by \citet{chu06,chu07}. These bubbles show ring structures with bright 8.0 $\mu$m emission. Enclosed by 8.0 $\mu$m shells, extended 24 $\mu$m emissions are generally confined to the inside of bubbles. \citet{chu06} proposed that these bubbles are primarily induced by hot young stars in massive star-forming regions. In an attempt to determine the nature of the bubbles detected by {\it Spitzer}, \citet{deh10} carried out an extensive study of 102 bubbles selected from the catalog of \citet{chu06} with MAGPIS \citep{hel06} 20 cm radio continuum data. They found that 86\% of the 102 bubbles contain ionized gas detected at 20 cm. The Milky Way Project, a citizen science initiative with more than 35,000 volunteers involved, has increased the number of known bubbles to more than 5,000 \citep{sim12}.

    \citet{tho12} found a statistically significant overdensity of massive young stellar objects (MYSOs) on borders of bubbles and speculated that triggered star formation plays an important role in the birth of massive stars in the Milky Way. And this tentative conclusion was further supported by the work of \citet{ken12}.

    Multi-wavelength studies have, in detail, revealed morphologies and star-forming activities in individual bubbles. Given the ability of providing velocity information, molecular lines are frequently observed to trace neutral material associated with bubbles \citep[e.g., in ][]{bea10,zha12,she12,ji12}. YSOs identified from cataloged point sources are used to investigate star-forming activities. Based on distributions of these YSOs and molecular structures, many investigators have claimed detections of triggered star formation in particularly studied bubbles \citep[e.g.,][]{kan09,bik10,dew12,ji12,zha13}.

    However, most of these well studied bubbles are relatively small ones. The kinematics and star formation properties of large bubbles with radii $>5^\prime$ have rarely been extensively investigated. It remains unclear whether these large bubbles are still under expansion, whether they can be regions of active star formation.

    In order to tentatively address these questions, we have selected a large infrared dust bubble (N6) from the catalog of \citet{chu06} to carry out a multi-wavelength study. Bubble N6 is located in the constellation of  Sagittarius with Galactic coordinates of $l=12\deg.512$ and $b=-0\deg.609$. This bubble has an average radius of 5.83 arcmin \citep{chu06}, which corresponds to a physical diameter of about 11 pc at a distance of 3.5 kpc (see Section \ref{sec:distance}). \citet{deh10} proposed that N6 is composed of two different structures: an open bubble in the northeast, and a filamentary ionization front in the southwest. In the interspace between these two structures, a bipolar system has been reported by \citet{yua12}. On the basis of observational analysis, \citet{yua12} posited that it would be a bipolar outflow driven by a massive protostar. However,  the physical status and the dynamic properties of the bubble N6 as a whole is still unclear.

    In this paper, we present observations of N6 in transitions of $^{12}$CO $J=1-0$ and $^{13}$CO $J=1-0$. Kinematics and molecular conditions are discussed in depth. With complementary survey data, we have revealed star-forming activities in this region. This paper is arranged as follows. We present a description of the observations and survey data in Section \ref{s-obs}, the results and some preliminary analysis in Section \ref{s-results}. In Section \ref{s-discussions}, we try to comprehensively discuss the data. We summarize our findings from this work in Section \ref{s-conclusions}.
\section{Observations and Data acquisition}\label{s-obs}
    \subsection{Molecular observations}

    Observations of the infrared dust bubble N6 in $J=1-0$ transitions of $^{12}$CO (115.272 GHz) and $^{13}$CO (110.201 GHz) ware carried out with the Purple Mountain Observatory (PMO) 13.7-m telescope on June 29$^{\rm th}$ 2012. The on-the-fly (OTF) mode was used to map a  $21\arcmin\times25\arcmin$ region with a reference point at $\alpha_{2000}={\rm 18^h14^m39^s.3}$, $\delta_{2000}=-18\deg24\arcmin42\arcsec.9$. The half power beam width (HPBW) of the PMO 13.7 m-telescope was about $52\arcsec$ at both 115 GHz and 110 GHz. The pointing and tracking accuracies were determined to be better than $5\arcsec$ by scanning planets (e.g., Jupiter and Venus).

    An SIS receiver \citep{sha12} with 9 beams was used to perform the observations. The maim beam efficiency at the center of the $3\times3$ array is about 0.44 at 115 GHz and 0.48 at 110 GHz. The efficiency of the outlying beams is slightly lower than that of the central one. The correction of beam efficiencies was performed during the observations. Mixers with local oscillators working at about 112.6 GHz make it possible for the 9 beams to receive signals from both lines simultaneously with $^{12}$CO $J=1-0$ and $^{13}$CO $J=1-0$ located in the upper sideband (USB) and lower sideband (LSB) with a 5 GHz separation. Each sideband has a width of 1 GHz. Eighteen fast Fourier transform spectrometers (FFTS) were used to digitize the signals and allocate 1 GHz to 16384 channels. This resulted in a spectral resolution of 61 kHz, corresponding to velocity resolutions of 0.16 km s$^{-1}$ and  0.17 km s$^{-1}$ at 115 GHz and 110 GHz, respectively. For the calibration of the intensity, the chopper-wheel method \citep{uli76} was used.

     The cloudy weather condition during our observations led to system temperatures high to about 550 K and 350 K at 115 GHz and 110 GHz. This resulted in rms noises of 1.7 K and 1.2 K in the brightness temperature for $^{12}$CO $J=1-0$ and $^{13}$CO $J=1-0$. Such high noise would make relatively weak signals undetectable. Nevertheless, regions with strong line emission can be validly probed. The velocity information provided by these data convincingly reveal the kinematics of the bubble and molecular conditions in some subregions.

     The data were reduced and visualized with the CLASS and GREG programs of the GILDAS package \citep{pet05}. Baselines were fitted with one order polynomial and removed for each line.

    \subsection{Survey Data}\label{sec-data-survey}
    Public survey data at wavelengths ranging from infrared to centimeter were retrieved from GLIMPSE, MIPSGAL, and MAGPIS surveys.

    GLIMPSE \citep[Galactic Legacy Infared Mid-Plane Survey Extraodinaire,][]{ben03,chu09} is a {\it Space Infrared Telescope Facility (SIRTF)} Legacy Science Program carried out using the InfraRed Array Camera \citep[IRAC,][]{faz04} onboard the {\it Spitzer Space Telescope} (SST) \citep{wer04}. The four wavebands of this instrument are centered at 3.6, 4.5, 5.8 and 8.0 $\mu$m with spatial resolutions ranging from $1.6\arcsec$ to $1.9\arcsec$. GLIMPSE surveyed the inner 130 degrees of the Galactic Plane and Galactic latitudes from $-1\deg$ to $1\deg$. The 5 $\sigma$ sensitivities of the four bands are 0.2, 0.2, 0.4, 0.4 mJy, respectively. In addition to images, the GLIMPSE survey performed point-source photometry using point spread functions (PSFs). The limiting magnitudes are 14, 12, 10.5, and 9.0 for bands 1 to 4 with photometric accuracy no worse than 0.2 mag. Photometric data in the \emph{J} (1.25 $\mu$m), \emph{H} (1.65 $\mu$m), and \emph{Ks} (2.17 $\mu$m) bands from the Two Micron All Sky Point Source Catalog \citep[2MASS PSC, ][]{skr06} are provided in the GLIMPSE Catalog to build-up a 7-bands photometric system. Both the images and the catalog of GLIMPSE are publicly available at the InfraRed Science Archive (IRSA)\footnotemark[1], where we have retrieved cutouts in the four IRAC bands and point sources in a $30\arcmin\times30\arcmin$ region centered at $\alpha_{2000}={\rm 18^h14^m47^s.09}$, $\delta_{2000}=-18\deg26\arcmin21\arcsec.2$. We restricted the extracted catalog to be a more reliable data set with the following criteria: a) Only sources with photometric errors no larger than 0.2 mag at 3.6 $\mu$m and 4.5 $\mu$m are taken into account; b) Photometric data with uncertainties larger than 0.2 mag at 5.6 and 8.0 $\mu$m are omitted; c) For the 2MASS bands, a threshold of 0.1 mag photometric error is used to forsake unreliable photometric values.

    \footnotetext[1]{http://irsa.ipac.caltech.edu/index.html}

    MIPSGAL \citep{car09} is a complement to the GLIMPSE legacy survey. This survey using the MIPS \citep[Multiband Infrared Photometer for {\it Spitzer},][]{rie04} instrument onboard the {\it Spitzer Space Telescope} (SST) surveyed an area comparable to that of GLIMPSE. The version 3.0 of MIPSGAL data includes mosaics (post basic calibrated data products) at 24 $\mu$m with sky coverage of $|b|<1\deg$ for $-68\deg<l<69\deg$, and $|b|<3\deg$ for $-8\deg<l<9\deg$. The spatial resolution and $5\sigma$ sensitivity at 24 $\mu$m are $6\arcsec$ and 1.7 mJy. From the IRSA server, we have extracted a cutout of a region same as that of GLIMPSE cutouts.

    We conducted point-source extraction and photometry of point sources in the MIPS 24 \micron~image using the PSF fitting capability of IRAF/DAOPHOT \citep{ste87}. The PSF was determined to be about 6\arcsec~by fitting the profiles of eight bright point sources in the field of interest. The standard deviation ($\sigma$) of the sky was estimated to be about $8.8\times10^{-6}~\mathrm{Jy~pixel^{-1}}$. \textsc{daofind} was used to extract point source candidates with a threshold of 20 $\sigma$. The final sources were determined by visually inspecting the extracted sources using \textsc{tvmark}. Sources affected by ghosts, diffraction spikes, halos from bright sources and artifacts residing in bright extended emission were rejected. Magnitudes of the extracted sources were determined using the magnitude zero point of 7.17 Jy provided in the MIPS instrument handbook\footnotemark[2]. Finally, we reached a catalog of 517 point sources with 24 \micron~photometry. Among them, 491 have IRAC counterparts within 2\arcsec.
    \footnotetext[2]{http://irsa.ipac.caltech.edu/data/SPITZER/docs/mips/mipsinstrumenthandbook/}

    We estimated the completeness of our catalog by counting the number of sources as a function of magnitude. A histogram plot of source magnitudes at each band has been plotted and carefully inspected. The magnitude, at which the number of sources drops sharply, has been treated as the complete limit. Finally, we found that our catalog is relatively complete to a magnitude of 13.5 at 3.6 \micron, 13.5 at 4.5 \micron, 12.2 at 5.8 \micron, 11.5 at 8.0 \micron, and 7.5 at 24 \micron.

    The Multi-Array Galactic Plane Imaging Survey \citep[MAGPIS,][]{hel06}, on its website\footnotemark[3], provides 90 cm images of the first Galactic quadrant. These images were obtained using the VLA at 325 MHz. The archived 90 cm data covering N6 have a spatial resolution of $25\arcsec$ and $1\ \sigma$ sensitivity of 5 mJy beam$^{-1}$ \citep{bro05}. A cutout around N6 has been extracted and used to reveal the large scale distribution of ionized gas.

    \footnotetext[3]{http://third.ucllnl.org/gps/}

\section{Results}\label{s-results}
     \begin{figure*}
    \centering
    \includegraphics[width=0.85\textwidth]{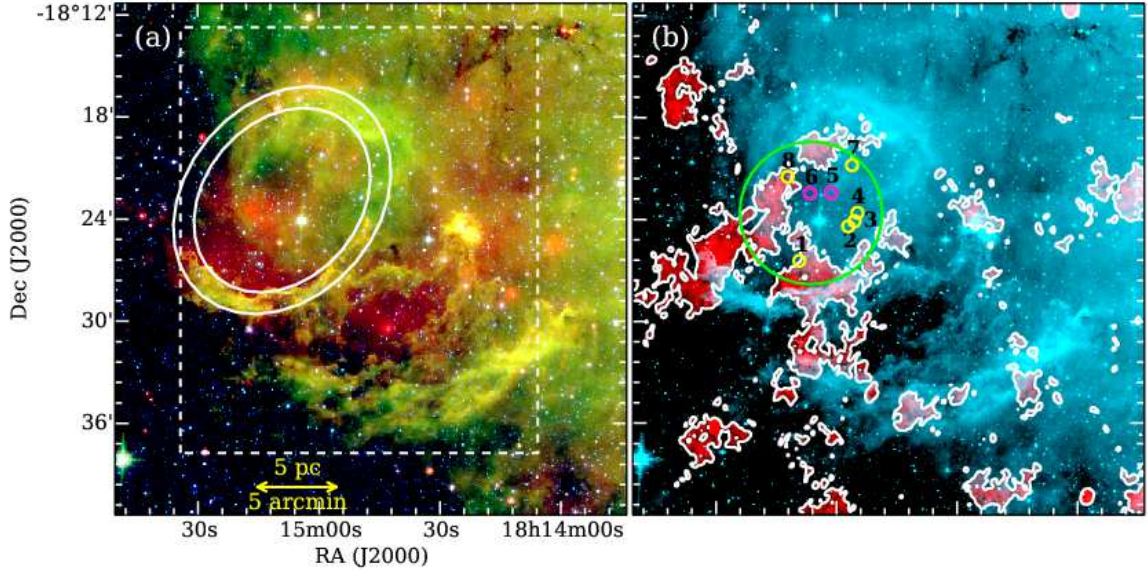}
    \caption{Composite color images of N6. {\it (a)} Compiled with {\it Spitzer} 4.5 $\mu$m (blue), 8.0 $\mu$m (green), and 24 $\mu$m (red) cutouts. The two ellipses represent the outer and inner rings of bubble N6. The major and minor radii of the inner ring are 5.94 and 4.58 arcmin, respectively. The outer ring has major and minor radii of 7.33 and 5.65 arcmin. The dashed box marks the region of PMO observations. {\it (b)} Compiled with {\it Spitzer} 8.0 \micron\ (cyan), and MGAPIS 90 cm (red) cutouts. The circle is used to confine a region, where candidate exciting stars reside. The small circles mark the potential exciting stars identified in Section \ref{sec:exciting}. \label{fig:composite}}
    \end{figure*}

    Shown in Figure \ref{fig:composite} are composite color images of N6 compiled using {\it Spitzer} and MAGPIS cutouts. As described in \citet{deh10}, N6 is composed of a ring in the northeast and a filamentary structure in the southwest. Both of these two components show bright emission at 8.0 $\micron$. The southwest bright 8 \micron~structure also could be a part of a large bubble or resulting from the interaction between ionizing radiations escaping from the holes of the main bubble and the the natal molecular cloud. However, it is too complicated to make clear the nature of this structure and beyond the scope of this work. In the following analysis, we mainly focus on the ring structure and its near vicinities. In Figure \ref{fig:composite}(a), a dashed box indicates the region of the PMO observations. In this section, we will show the results of our PMO observations and the identification of YSO candidates.

    \subsection{Molecular Clouds} \label{sec:mole}

    We have carried out observations of a $21^\prime\times25^\prime$ region in the $J=1-0$ transitions of \co\ and \tco. These data have provided information of molecular cloud and velocity distributions in bubble N6.

    \subsubsection{Four velocity Components}
    \begin{figure}
    \centering
    \includegraphics[width=0.45\textwidth]{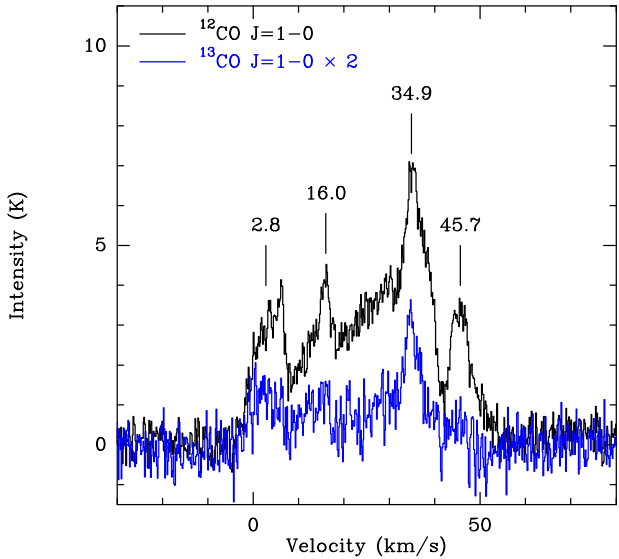}
    \caption{Averaged spectra of lines of $^{12}$CO $J=1-0$ and $^{13}$CO $J=1-0$. We note that the intensity of $^{13}$CO $J=1-0$ is multiplied by 2 for clarity. \label{fig:spec-aver}}
    \end{figure}

    \begin{figure*}
    \centering
    \includegraphics[width=\textwidth]{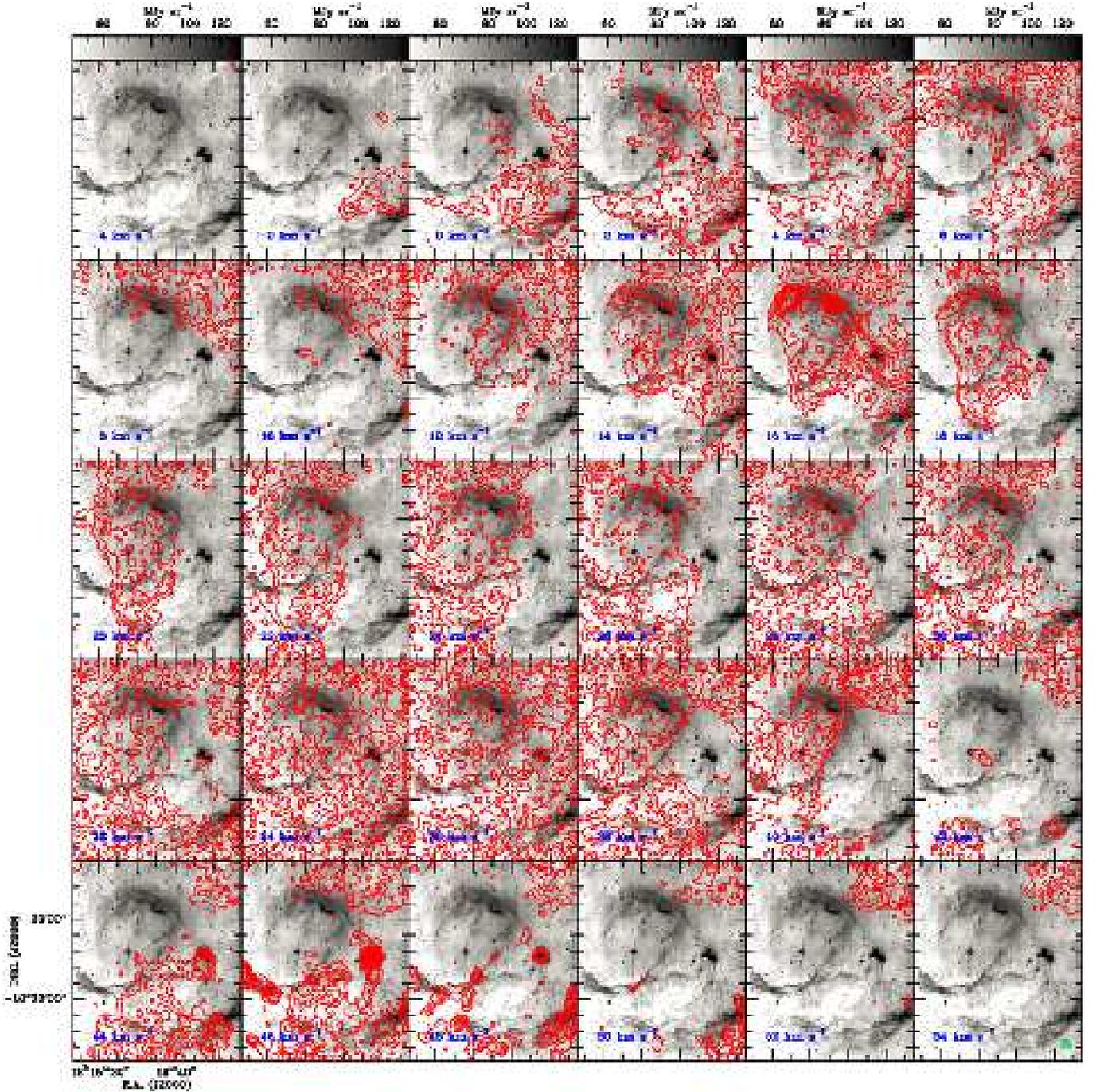}
    \caption{Channel maps of $^{12}$CO $J=1-0$ (contours) overlayed on the {\it Spitzer} 8.0 \micron\ image. The velocity interval of integration for each small panel covers 2 km s$^{-1}$ with the center value marked in the left-bottom corner of each panel. The contours start from 5 $\sigma$ with steps of 2.5 $\sigma$ ($1\ \sigma=2$ K km s$^{-1}$). In the right-bottom corner of the last panel, a small circle is used to present the primary beam of our PMO observations. \label{fig:channel-map} }
    \end{figure*}

    \begin{figure*}
    \centering
    \includegraphics[width=0.9\textwidth]{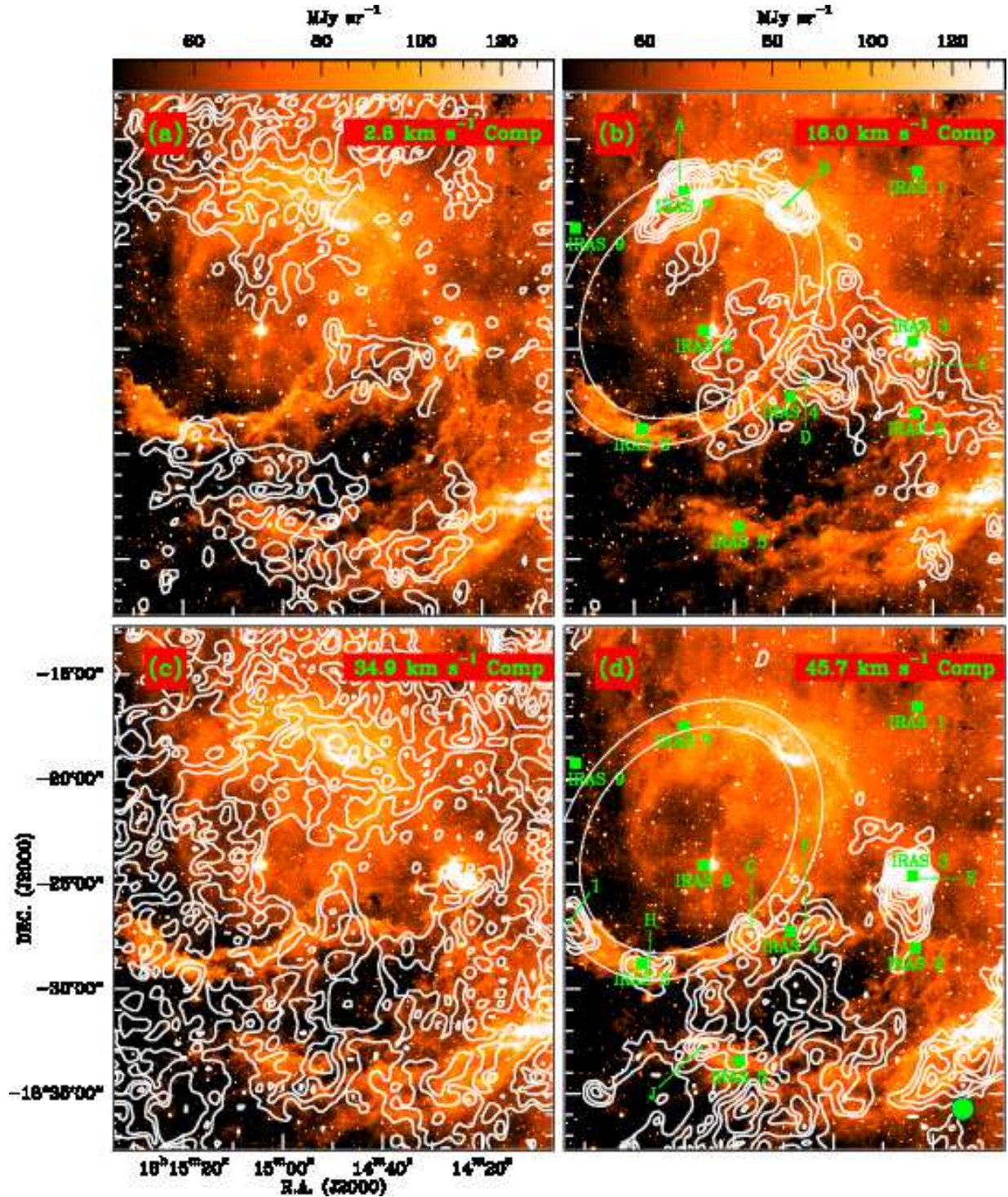}
    \caption{Velocity integrated intensities (contours) of the four components of $^{12}$CO $J=1-0$ overlayed on IRAC 8.0 $\mu$m cutout. Velocity intervals of integrations are [1, 9], [14, 18], [33, 49], and [43, 49] for the 2.8, 16.0, 34.9, and 45.7 components, respectively. The contour levels start from 4 $\sigma$ with steps of 1.5 $\sigma$. Here, 1 $\sigma$ noises for the four components are 3.5, 2.5, 4.2, and 4.0 K km s$^{-1}$, respectively. The identified molecular clumps are indicated with A-J in the right panels. The small filled squares are representative of IRAS sources. The filled circle in the right-bottom corner of the 45.7 km s$^{-1}$ component panel marks the primary beam of our PMO observations. \label{fig:co-irac} }
    \end{figure*}

    \begin{figure*}[t]
    \centering
    \includegraphics[width=0.95\textwidth]{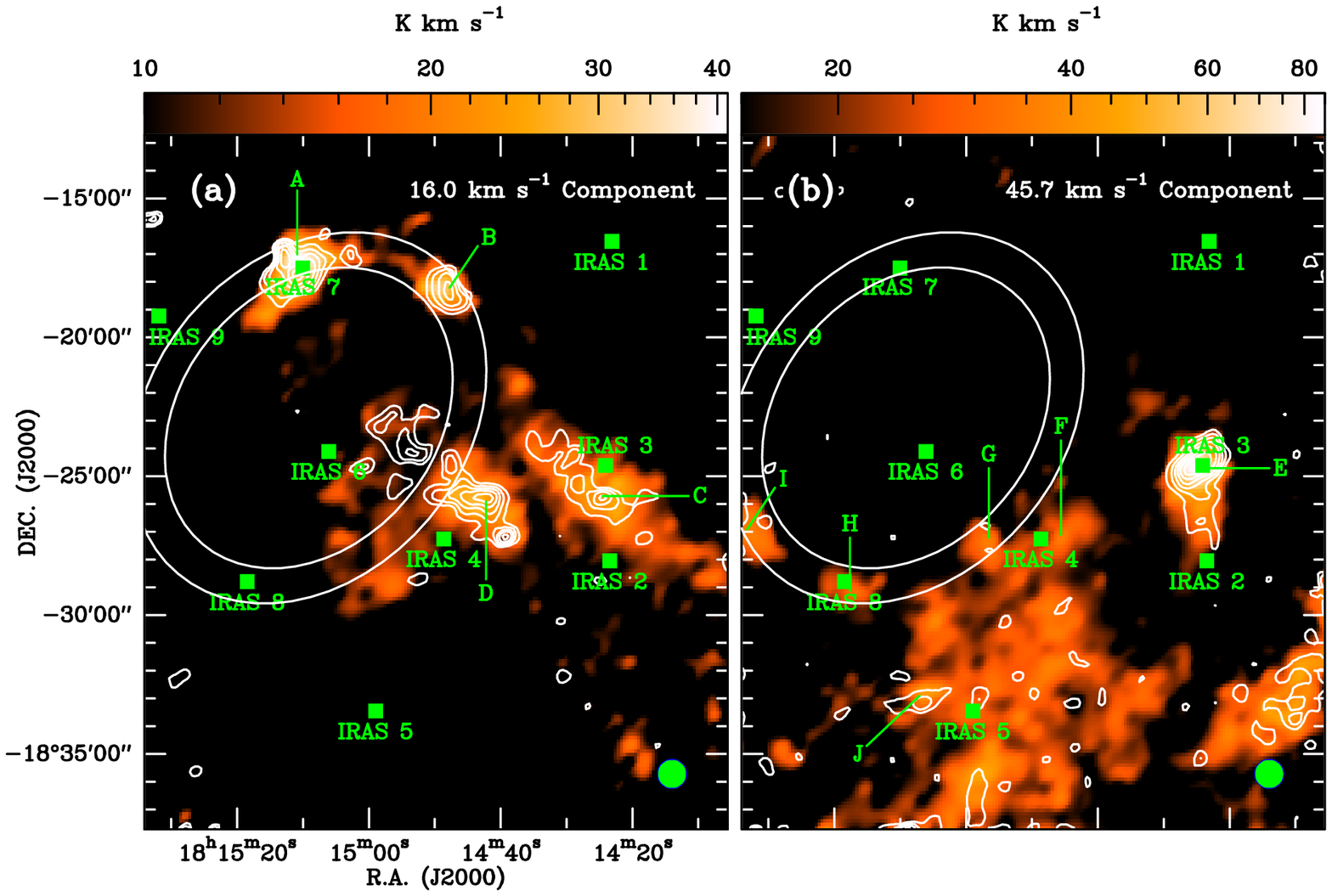}
    \caption{Velocity integrated intensities of the 16.0 km s$^{-1}$ component ({\it a}) and 45.7 km s$^{-1}$ component ({\it b}) of $^{13}$CO $J=1-0$ overlayed on that of $^{12}$CO $J=1-0$. The velocity intervals of integrations of the 16.0 and 45.7 km s$^{-1}$ components of \tco\ $J=1-0$ are [14,18] and [42,49] km s$^{-1}$, respectively. The contours start from 4 $\sigma$ with steps of 1.3 $\sigma$. Here, 1 $\sigma$ noises are 1.3 and 3.0 K km s$^{-1}$ for the 16.0 and 45.7 km s$^{-1}$ components. The symbols are the same as that in Figure \ref{fig:co-irac}.} \label{fig:co-13co}
    \end{figure*}

    Molecular conditions of N6 are extraordinarily complicated. The spectra of both lines of \co\ and \tco\ show more than one velocity components. On the basis of the averaged spectra of the whole observed region (see Figure \ref{fig:spec-aver}), we have identified four velocity components centered at 2.8, 16.0, 34.9, and 45.7 km s$^{-1}$, respectively. In order to unambiguously ascertain the relationship of each velocity component with N6, we have constructed channel maps of \co\ $J=1-0$, which are shown in Figure \ref{fig:channel-map}. The \co\ emissions show diffuse features at the majority of velocity channels. However, several molecular condensations are detected at velocities of 13 to 19 km s$^{-1}$ and 43 to 49 km s$^{-1}$. Most of them are spatially associated with the ring and the filamentary structure, suggesting physical correlation with N6.

    In Figure \ref{fig:co-irac}, we have overlayed the velocity integrated intensity as contours of each velocity component on the {\it Spitzer} 8.0 \micron\ image. The emissions of these four velocity components show different features.

    \begin{itemize}
      \item The 2.8 km s$^{-1}$ component is dominantly detected on the north and south sides (see Figure \ref{fig:co-irac}(a)). Molecular material at this velocity is diffusely distributed. Compared with the 16.0 and 45.7 km s$^{-1}$ components, the material at this velocity seems not correlated with bubble N6. Using the Galactic rotation curve in \citet{rei14}, we have calculated the near and far kinematic distances to this velocity component. The near distance of 0.4 kpc indicates that the 2.8 km s$^{-1}$ velocity component may trace a foreground region residing in the Sagittarius arm.
      \item As shown Figure \ref{fig:co-irac}(b)\&(d), several molecular condensations are conspicuous in the 16.0 and 45.7 km s$^{-1}$ components. Five of them are detected on the border of the ring structure. This suggests that these two velocity components are associated with N6. The difference of the velocities could be due to the expansion of the ring.
      \item Molecular gas at the 34.9 km s$^{-1}$ component is diffusely distributed throughout the whole mapped region. As the velocity is in-between the two components associated with N6, this component may originate from the natal cloud where the exciting star(s) of N6 have formed. There is a void in the middle of the integrated intensity map. It is spatially consistent with the cavity detected at MAGPIS 90 cm. The emergence of this void (or cavity) would be attributed to winds from the the exciting star(s).
    \end{itemize}

    Toward N6, a very broad recombination line at 9 cm was reported by \citet{loc89}. The recombination line is peaked at 40 km s$^{-1}$ with a FWHM width of 33.2 km s$^{-1}$. The full width exceeds 60 km s$^{-1}$, having the 16.0, 34.9, and 45.7 km s$^{-1}$ components revealed by \co\ covered by the recombination line. This provides additional evidence to the consistence of the 16.0, 34.9, and 45.7 km s$^{-1}$ components with N6.

    To conclude, we posit that the 16.0 and  45.7 km s$^{-1}$ components originate from the expansion of the ring of N6, and the 34.9 km s$^{-1}$ component is tracing the natal cloud. In the following, we estimate the distance to N6 using a systemic velocity of 34.9 km s$^{-1}$.

    \subsubsection{Distance Estimation}\label{sec:distance}

    The distance to N6 has not been well confined. Based on a recombination line velocity of 40 km s$^{-1}$ \citep{loc89}, \citet{deh10} proposed a near and a far kinematic distances of 4.12 and 12.49 kpc. The velocity resolution in \citet{loc89} is about 4 km s$^{-1}$. Such poor value would induce large uncertainties to the estimated distances. Our observations have a velocity resolution of 0.17 km s$^{-1}$, an order of magnitude higher than that in \citet{loc89}. Based on our observations and aforementioned analysis, we suggest a more reliable systemic velocity of 34.9 km s$^{-1}$. Using the Galactic rotation curve of \citet{rei14}, we have obtained the near and far kinematic distances to be 3.5 and 12.8 kpc.

    Previous works have suggested that most star-forming regions are located on one of the Galactic spiral arms \citep{kol03,rom09}. To resolve the near-far kinematic distance ambiguity, we have checked the location of N6 on the face-on plane of our Galaxy. The four arms model of \citet{val05} has been used. We noticed that N6 is located on the Scutum arm while the near distance of 3.5 kpc is employed and the far distance of 12.8 kpc would put N6 on a location in between the Sagittarius and Perseus arms. Thus, we have adopted the near value of 3.5 kpc as the distance to N6.

    Kinematic distances always have relatively large uncertainties. The origins of such uncertainties could come from the inaccuracies of systemic velocities and fluctuations of rotation curves in use. On the basis of investigations of Galactic \hii~regions, \citet{and12} proposed that the uncertainties of kinematic distances could change from smaller than 5\% to larger than 20\%. The uncertainties could be small for sources on the Galactic plane, and relatively large for source with high latitudes. Given that N6 resides in the inner disk, we adopt a 10\% uncertainty for the kinematic distance in the following analysis.

    \subsubsection{Clumps}\label{sec:clumps}

    In the velocity integrated intensity maps of 16.0 km s$^{-1}$ and 45.7 km s$^{-1}$ components of \co\ $J=1-0$, we have identified ten peaks as molecular condensations. Due to the relatively poor spatial resolution of our observations, the sizes of these condensations would be larger than 0.9 pc (the beam size of our PMO observations). Thus we refer to these sources as ``clumps", which have sizes of several tenths to a few parsecs and could serve as forming sites of clusters\citep{wil00,ber07}.   We have labeled them as A-J in Figures \ref{fig:co-irac}\&\ref{fig:co-13co} and tabulated them in Table \ref{tb-cl-ob}. Four of these clumps are coincident with the 16.0 km s$^{-1}$ component. The other six are detected in the 45.7 km s$^{-1}$ component. Five clumps (i.e., A, B, G, H, and I) are distributed on the border of the ring structure. The clump E spatially coincides with the bipolar structure reported in \citet{yua12}.

    For identifying local peaks as clumps, a threshold of 8 $\sigma$ (about 32 K km$^{-1}$) was used. There are additional local peaks in the right panels of Figure \ref{fig:co-irac}, and could also be clump candidates. We carefully inspected the \co~$J=1-0$ spectra, and found they are fake peaks whose emersion could be attributed to fluctuations of baselines and worse sensitivity in the edge of the mapped region.

    Figure \ref{fig:co-13co} presents velocity integrated intensity maps of 16.0 km s$^{-1}$ and 45.7 km s$^{-1}$ components of \tco\ $J=1-0$ overlayed on that of \co\ $J=1-0$. Only five clumps have detections in \tco\ $J=1-0$. They are clumps A-E.

    For each identified clump, we have fitted a two-dimensional Gaussian to the area around the peak to determine the size. For the five clumps with \tco\ detections, the fittings have been performed based on integrated intensity maps of \tco\ $J=1-0$. For the other five ones, integrated intensity maps of \co\ $J=1-0$ have been used. The resulting major and minor axes and position angle for each source are given in Table \ref{tb-cl-ob}. The effective radius of each clump is determined via the relation of $R=\sqrt{D_{major}\cdot D_{minor}}/2$ and given in the column 2 of Table \ref{tb-cl-der}. $D_{major}$ and $D_{minor}$ are the deconvolved major and minor axes of the 2-D Gaussian, respectively.

    We have carried out one-component Gaussian fittings to line profiles at peaks of the clumps. This procedure has resulted in integrated intensities, brightness temperatures, velocity centroids, and full widths at half-maximum (FWHMs) of \co\ $J=1-0$ and \tco\ $J=1-0$. The resulting parameters are given in Table \ref{tb-cl-ob}.

    We have followed a rotation temperature-column density analysis to estimate local thermodynamic equilibrium (LTE) parameters of the clumps with reliable detection in both lines. The residual brightness temperature ($T_r$) for a specific transition as a function of the excitation temperature ($T_{ex}$) can be expressed as
    \begin{equation}
    T_{r}=\frac{h\nu}{\kappa}[\frac{1}{\mathrm{exp}(\frac{h\nu}{\kappa T_{ex} })-1}-\frac{1}{\mathrm{exp }(\frac{h\nu}{\kappa T_{bg}})-1}]\times[1-\mathrm{exp }(-\tau)]f, \label{eq-Tex}
    \end{equation}
    where $T_{\rm bg}=2.73$ K is the temperature of the cosmic background radiation, and $f$ is the beam-filling factor. Optical depths of both \co\ and \tco\ can be directly obtained from comparing the measured brightness temperatures \citep{gar91}.
    \begin{equation}
    \frac{T_r{\rm (^{12}CO)}}{T_r{\rm (^{13)}CO}}\approx\frac{1-{\rm exp} (-\tau_{12})}{1-{\rm exp} (-\tau_{13})}. \label{eq-tau}
    \end{equation}
    In the procedure of estimating the distance, a Galactocentric radius of 4.9 kpc was obtain for N6. This value locates this bubble in the 5 kpc molecular ring of our Galaxy. Thus an isotope ratio of ${\rm [^{12}CO]/[^{13}CO]}=\tau_{12}/\tau_{13}=53$ \citep{wil94} is applied to derive the optical depths which are given in column 3 and 4 of Table \ref{tb-cl-der}. With an assumption of a beam-filling factor of 1, the excitation temperature of each clump can be straightforwardly obtained and is presented in column 5 of Table \ref{tb-cl-der}. The resulting excitation temperatures range from 16 K to 30 K, consistent with that in intermediate- to high-mass star-forming regions \citep{mol96,sri02,wu06,liu11,ren14}.

    According to \citet{gar91}, the column density of a linear molecule can be expressed as
    \begin{eqnarray}
    N=\frac{3\kappa}{8\pi^3B\mu^2}\frac{\mathrm{exp }[hBJ(J+1)/\kappa T_{ex}]}{(J+1)} \nonumber \\
    \times\frac{(T_{ex}+hB/3\kappa)}{[1-\mathrm{exp }(-h\nu/\kappa T_{ex})]}\int\tau dv. \label{eq-lte-density}
    \end{eqnarray}
    where $B$ and $\mu$ are the rotational constant and permanent dipole of the molecule, and $J$ is the rotational quantum number of the lower state of the observed transition. We have calculated the \tco\ column density in the peak of each clump. To obtain the molecular hydrogen column density, we have adopted a canonical [CO]/[H$_2$] abundance ratio of $\simeq10^4$ and an isotope ratio of ${\rm [^{12}CO]/[^{13}CO]}=53\pm4$ \citep{wil94}. Then, the mass of a clump has been resulted from $M_{LTE}=(4/3)\pi R^3n_{{\rm H_2}}\mu_gm({\rm H_2})$. Here, $\mu_g=1.36$ is the mean atomic weight of gas, $m({\rm H_2})$ is the mass of a hydrogen molecule, and $n_{{\rm H_2}}=N_{\rm H_2}/2R$ is the volume density of molecular hydrogen. The resulting densities and masses are given in columns 6-8 of Table \ref{tb-cl-der}. The five clumps, with reliable detection in both lines, have masses from several hundred to high to more than 5,000 $M_\sun$. This corroborates that they could serve as cradles for massive stars.

    \begin{figure}
    \centering
    \includegraphics[width=0.5\textwidth]{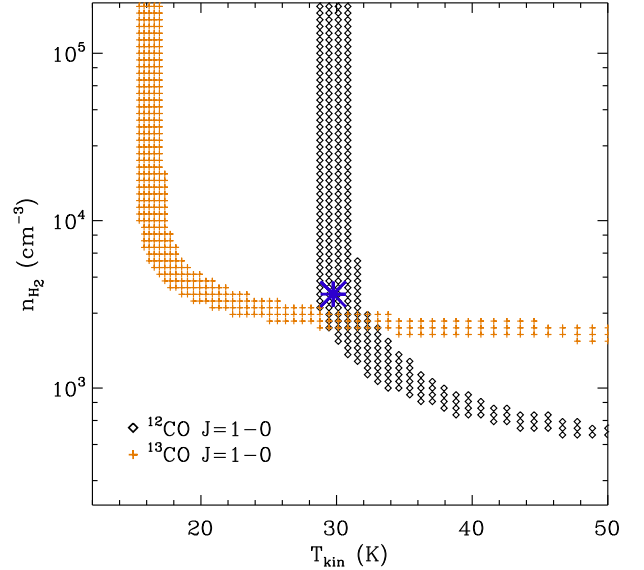}
    \caption{Non-LTE model fitting of $^{12}$CO $J=1-0$ and $^{13}$CO $J=1-0$ at the emission peak of clump E. The asterisk is representative of the LTE result. Here, we only illustratively present the plot of clump E. \label{fig:radex}}
    \end{figure}

    A computer program (\verb"RADEX") for the fast non-LTE analysis of interstellar line spectra has been developed by \citet{van07}. With collisional and radiative processes involved in the code, \verb"RADEX" can be used to effectively constrain the densities and temperatures of a clump in which two molecular transitions have been positively detected. We have followed \citet{liu12} to explore a grid of H$_2$ volume densities and kinetic temperatures in ranges of $10^2$-$10^6$ cm$^{-3}$ and 5-50 K. Expanding spherical geometries have been used. Models fulfilling the restriction of $|T_{mod}-T_r|\leq\sigma$ are preserved for further analysis. The $T_{mod}$ and $T_r$ are modeled and measured brightness temperatures, respectively. These procedures have been performed for the \co\ $J=1-0$ and \tco\ $J=1-0$ transitions in the peaks of clumps A-E. The final acceptable parameters have been reached via comparing models of \co\ transition to that of \tco\ transition. Shown in Figure \ref{fig:radex} is the modeled H$_2$ volume density as a function of kinetic temperature. In this paper, we just illustratively present the plot of clump E. The valid models would be the ones overlapped in Figure \ref{fig:radex}. Averaged parameters from the valid models of each clump are given in columns 9-14 of Table \ref{tb-cl-der}. We note that parameters (e.g., temperatures, densities, and masses) obtained from the LTE and non-LTE approaches are well consistent with each other. The excitation temperatures from the non-LTE estimation are slightly lower than that from the LTE calculation, and also smaller than the kinetic temperatures. We get similar LTE and non-LTE masses for clumps B, C, and D.

    The lack of reliable detections in \tco\ $J=1-0$ has prevented us estimating the LTE parameters of clumps F-J. However, some non-LTE parameters can be obtained using the \verb"RADEX" program with an assumed kinetic temperature of 20 K. We have used the \textit{Python} script provided by \citet{van07} to calculate \co\ column densities. The H$_2$ column density has been reached by using a canonical abundance ratio of ${\rm [^{12}CO]/[H_2]=10^{-4}}$. The resulting column density of each clump is provided in Table \ref{tb-cl-der}, in which the non-LTE volume density and mass are also given. In contrast to that of clumps with detections in \tco\ $J=1-0$, the densities and masses of clumps F-J are relatively small.

    \subsection{IRAS Point Sources}

    We queried the V2.1 version of the \textit{IRAS} point source catalog\citep{bei88} and found that there are nine \textit{IRAS} point sources in the mapping region of our PMO observations. They are labeled as IRAS 1-9 with ascending right ascension and indicated in Figures \ref{fig:co-irac} and \ref{fig:co-13co} with filled squares. Their genuine IRAS names, IRAS photometric data, and photometric quality flags from the catalog are given in Table \ref{tb-iras}. Noticeably, IRAS 18114-1825 (IRAS 3), IRAS 18122-1818 (IRAS 7), and IRAS 18123-1829 (IRAS 8) are associated with clumps E, A, and H, respectively.

   Infrared flux densities of the nine \textit{IRAS} sources can be obtained with the following equation \citep{cas86}:
    \begin{equation}
    F(10^{-13} \mathrm{W m^{-2}})=1.75\times(\frac{F_{12}}{0.79}+\frac{F_{25}}{2}+\frac{F_{60}}{3.9}+\frac{F_{100}}{9.9}).
    \end{equation}
   Here, $F_{12}$, $F_{25}$, $F_{60}$, and $F_{100}$ are flux densities in Jy at 12 $\mu$m, 25 $\mu$m, 60 $\mu$m, and 100 $\mu$m, respectively. With the distance of 3.5 kpc, we have obtained infrared luminosities, which are given in column 10 of Table \ref{tb-iras}. Color indices of log(${F_{25}}/{F_{12}}$) and log(${F_{60}}/{F_{12}}$) are also calculated and shown in Table \ref{tb-iras}.

    \subsection{YSO Candidates in N6}\label{sec:YSOs}

    Young stellar objects always show excessive infrared emission which can be effectively used for discriminating YSOs from field stars and distinguishing different evolutionary stages. At a considerably early evolutionary stage, protostars are mostly embedded in dust envelopes. They exhibit large excessive infrared emission and take on infrared spectral indices $\alpha_{\rm{IR}}>-0.3$ indicative of flat or ascending spectral energy distributions (SEDs) at wavelength longward of 2 $\mu$m \citep{lad87,gre94}. For pre-main sequence (PMS) stars with optically thick disks, the SEDs tend to descend and the infrared spectral indices are in the range of $-1.6<\alpha_{\rm{IR}}<-0.3$. ``Transition disk" (TD) sources are more evolved YSOs. The inner part of the disks have been cleared by photoevaporation of central stars or by planet forming processes. Such YSOs would be only excessively emitting at wavelengthes longer than 16 $\mu$m \citep{str89}.

    These properties of YSOs make photometric observations in the near- to mid-infrared plausible for identifying them from field stars. Color-based source identification and classification schemes have been developed and verified practical. In this subsection, we identify potential YSOs following the schemes proposed by \citet{gut09}. The resulting YSOs are classified into Class I (protostars, including Class 0, Class I, and ``flat spectrum" sources), Class II, and TD sources. These YSO candidates are further inspected by fitting their SEDs using the online tool of \citet{rob07}.

    \subsubsection{Contaminants Removal}
    There are several kinds of contaminants that would be misidentified as YSOs in our original sample. Extragalactic contaminations could stem from star-forming galaxies and broad-line active galactic nuclei (AGN), which show PAH-featured emission yielding very red 5.8 and 8.0 $\mu$m colors \citep{ste05,gut08}. \citet{gut09} developed an identification scheme based on color-color spaces of $[4.5]-[5.8]\ versus\ [5.8]-[8.0]$ and $[3.6]-[5.8]\ versus\ [4.5]-[8.0]$ to reject star-forming galaxies. They utilize the $[4.5]\ versus\ [4.5]-[8.0]$ color-magnitude diagram to flag broad-line AGNs. We follow their criteria to identify and eliminate these two kinds of contaminants from our sample. In our own Galaxy, unresolved knots of shock emission and resolved PAH emission are often detected in the IRAC bands, yielding additional contaminations \citep{gut09}. Based on color indices of $[3.6]-[4.5]$ and $[4.5]-[5.8]$ \citep[for details, please refer to][]{gut09}, these sources are weeded out.

    \subsubsection{Identification and Classification}

    \begin{figure}
    \centering
    \includegraphics[width=0.45\textwidth]{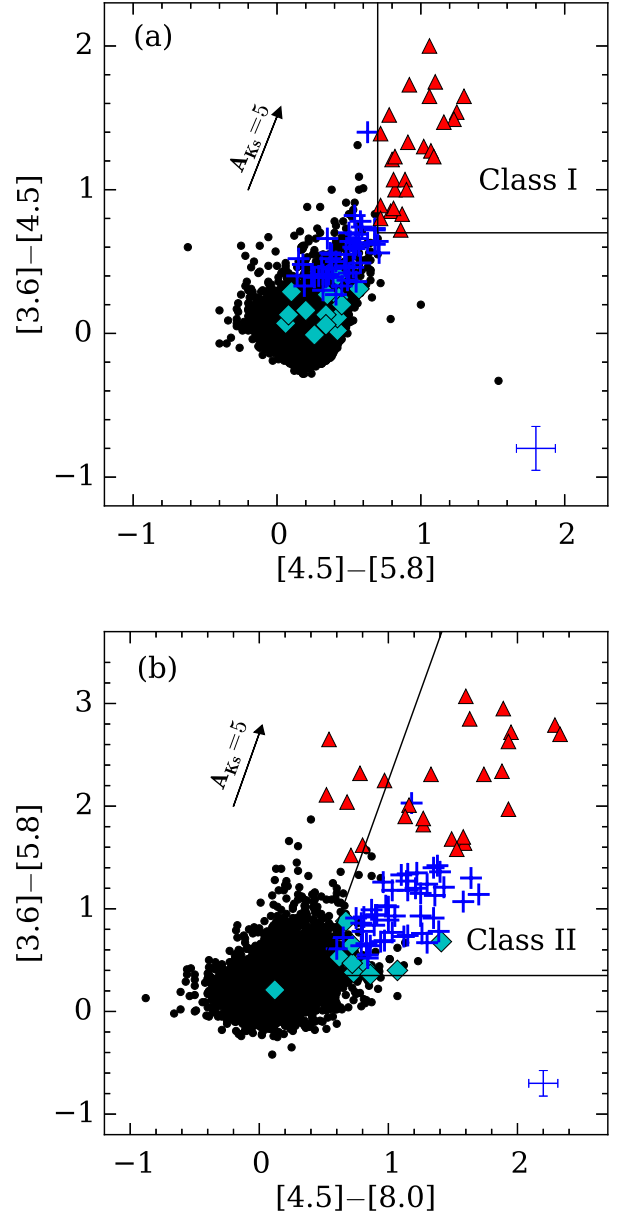}
    \caption{IRAC color-color diagrams showing the distribution of field stars (black dots), 'transition disk' (cyan diamonds), Class II (blue crosses), and Class I (red triangles) sources. The solid lines in the {\it (a)} and {\it (b)} panels indicate the defining loci of Class I and Class II sources \citep[for details, please refer to][]{gut09}. \label{fig:ccd-IRAC}}
    \end{figure}

    We have followed the source classification schemes developed by \citet{gut09} to identify and categorize potential YSOs from sources with  contaminants excluded.

    In the first phase, only sources with valid detections in all four IRAC bands have been considered. Any sources fulfilling color criteria of $[3.6]-[4.5]>0.7$ and $[4.5]-[5.8]>0.7$ are regarded as Class I YSOs \citep{gut09}.  In the remaining pool, Class II sources have been picked out based on the constraints of \emph{i}) $[3.6]-[4.5]-\sigma_1>0.15$, \emph{ii}) $[3.6]-[5.8]-\sigma_2>0.35$, \emph{iii})$[4.5]-[8.0]-\sigma_3>0.5$, and \emph{iv}) $[3.6]-[5.8]+\sigma_2\le\frac{0.14}{0.04}\times(([4.5]-[8.0]-\sigma_3)-0.5)+0.5$ \citep{gut09}. Here, $\sigma_1=\sigma([3.6]-[4.5])$, $\sigma_2=\sigma([3.6]-[5.8])$,  and $\sigma_3=\sigma([4.5]-[8.0])$ are combined errors, added in quadrature.

    \begin{figure}
    \centering
    \includegraphics[width=0.45\textwidth]{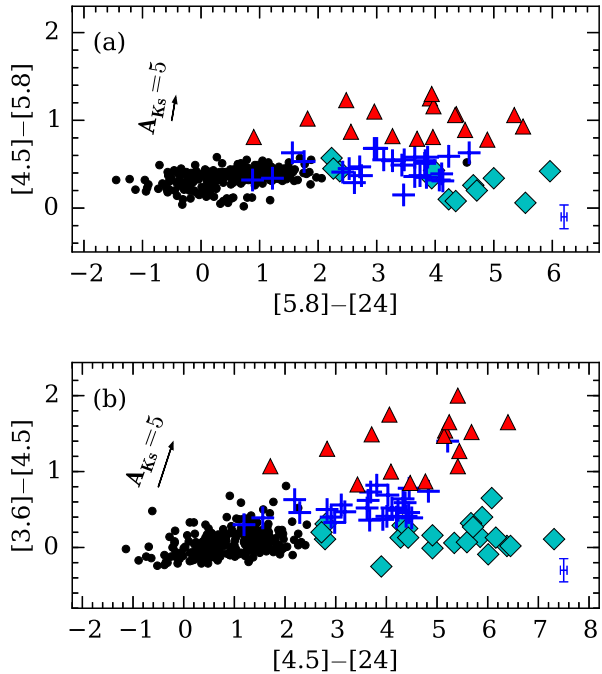}
    \caption{IRAC-MIPS color-color diagrams. Field stars, 'transition disk', Class II, and Class I sources are represented with black dots cyan diamonds, blue crosses, and red triangles, respectively.}\label{fig:ccd-mips}
    \end{figure}

    In the second phase, sources with 24 \micron~data but not classified to be YSOs in previous procedure are reexamined. Sources with colors of $[5.8]-[24]>2.5$ or $[4.5]-[24]>2.5$ are classified to be TDs \citep{gut09}. With the potential contaminants and YSOs identified above excluded, there still remain some sources with bright 24 \micron~emission. For sources lacking valid  photometric data in one or more of the IRAC bands, they are picked out as additional Class I type YSOs once fulfilling $[24]>7$ and $[X]-[24]>4.5$ mag, where $[X]$ is the longest wavelength IRAC detection that we have \citep{gut09}.

    The classification scheme of \citet{gut09} includes three phases. Compared with the methods mentioned above, they also use 2MASS data to identify additional YSOs. Considering that the distance of N6 is much farther than the sources in \citet{gut09}, 2MASS photometry would be severely affected by interstellar extinction. The use of 2MASS data to identify YSOs would induce heavy contamination from foreground field stars.

\begin{figure*}
    \centering
    \includegraphics[width=0.9\textwidth]{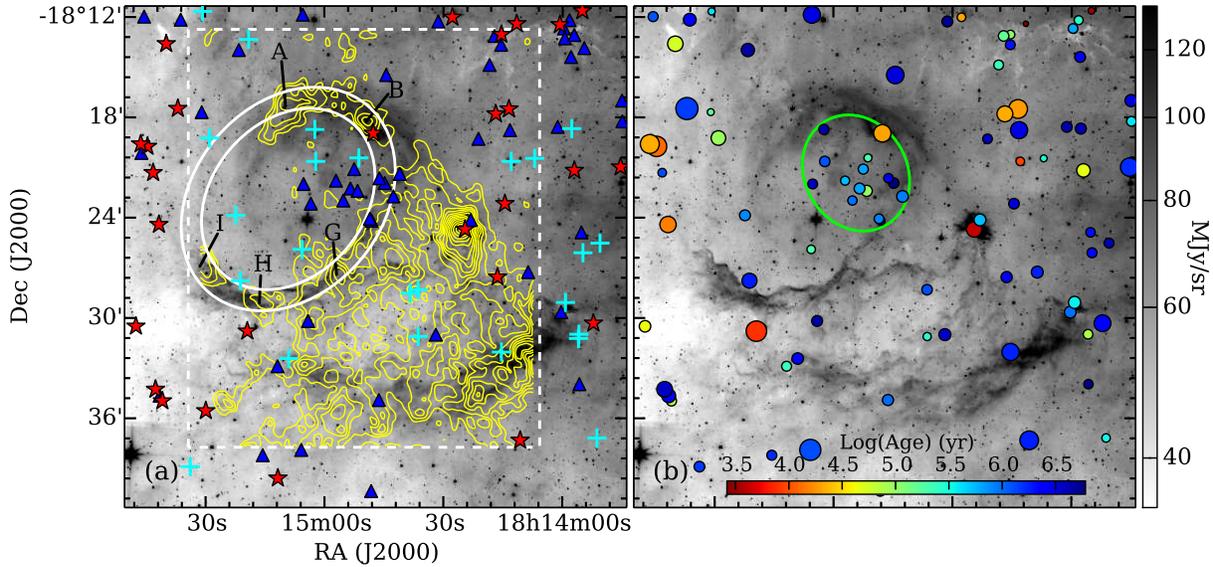}
    \caption{Spatial distribution of YSO candidates. The grayscale represents 8.0 \micron~emission. The contours show the combined velocity integrated intensity of the 16.0 and 45.7 km s$^{-1}$ components of  \co\ $J=1-0$. The velocity interval of integration covers [14, 18] and [43, 49] km s$^{-1}$. The contour levels start from $3\sigma$ ($1\sigma=5$ K km s$^{-1}$). The crosses, triangles, and stars represent 'transition disk', Class II, and Class I sources, respectively. The filled circles in pannel (\emph{b}) mark the 87 YSOs with valid SEDs. The sizes indicate the masses from 0.9 to 16 \msun. \label{fig:yso-irac} }
    \end{figure*}

   The above identification procedures have resulted in 99 YSOs which are classified into 27 Class I, 48 Class II, and 24 TD objects. A summary of the results is given in Talbe \ref{tb-count}. Their distributions on different color-color diagrams are shown in Figures \ref{fig:ccd-IRAC}\&\ref{fig:ccd-mips}. We note that the color based classification scheme would lead to misidentifications. A Class II YSO viewed at high inclination would show features resembling that of a Class I source. An edge-on Class I YSO can have similar infrared color of a Class 0 source. Thus, all Class I, II, and TD sources identified in this paper are candidates. The distribution of all the identified YSOs is shown in Figure \ref{fig:yso-irac}(a). Although most YSO candidates are distributed outside the bubble, there are a small group of YSOs residing inside the shell-like structure. Detailed discussions of star formation in this region is given in $\S$\ref{sec:SF}.

\begin{figure}
    \centering
    \includegraphics[width=0.45\textwidth]{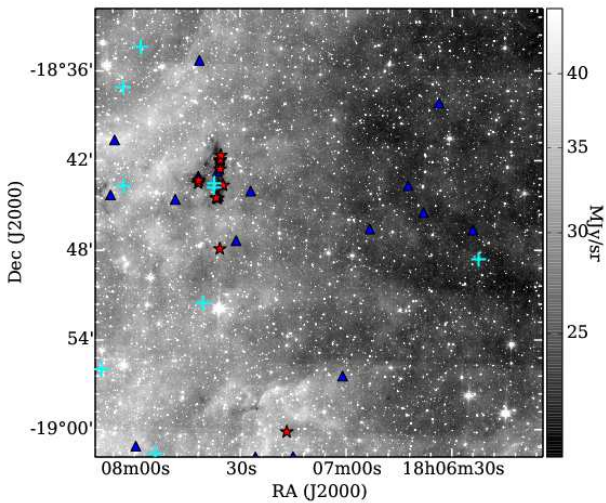}
    \caption{Spacial distribution of YSO candidates in the control field. The background and symbols are the same as that in Figure \ref{fig:yso-irac}. \label{fig:control} }
    \end{figure}

    In order to check the validation of the source classification scheme used in this work, we have tried to identify YSOs in a control field centered at ($l=11.2461$, $b=0.8646$) following the same procedures. The control field have been chosen based on the angle distance to N6 (as near as possible) and the intensity at 8.0 \micron~(no extended structure brighter than 50 MJy sr$^{-1}$). In a region of the same area of the target field, 39 YSO candidates have been identified. The results are outlined in Table \ref{tb-count}. The distribution of all the identified YSOs in the control field is shown in Figure \ref{fig:control}. Interestingly, 15 of the 39 YSOs concentrate at the east to form a small cluster which is spatially in coincidence with a small infrared dark filament. They could be real YSOs with high possibilities. If the other 24 sources were treated as potential misidentifications in a $30^\prime\times30^\prime$ region, we can estimate that more than 75\% of the 99 YSOs in the N6 field are with high reliability.

    \subsubsection{SED Fitting}

    We have fitted the SEDs of all the 99 YSO candidates using the online tool of \citet{rob07}. The SED fitting tool was developed based on a grid of 20,000 two-dimensional Monte Carlo radiation transfer models spanning a large range of stellar masses and evolutionary stages. Each model consists of parameters of a central star, a flared accretion disk, a rotationally flattened envelope, and bipolar cavities. There are outputs at 10 viewing angles (inclinations) for each model. Thus, the fitting tool has 200,000 SEDs to choose from. Based on a regression algorithm, SEDs within a specific $\chi^2$ are returned.

    The required inputs are photometric data at multi-bands, the distance and interstellar extinction ranges for fitting SEDs of a source. In this work, photometric data at 2MASS and IRAC bands are from the GLIMPSE catalog, while the 24 \micron~data were extracted from the 24 \micron~image using the IRAF/DAOPHOT (see $\S$\ref{sec-data-survey}). Distances are constrained in the range of 3-4 kpc. The interstellar extinction towards N6 has been estimated to be about 21 mag based on the extinction model S (with spiral arms) of \citet{amo05}. Thus, we have constrained the extinctions for all the 99 YSOs in the range of 0-30 mag with a relatively large relaxation.

    There are more than one modeled SEDs returned for each source. The one with the smallest $\chi^2$ ($\chi_\mathrm{best}^2$) is the best fitted SED. However, it does not necessarily represent the true natures of the target. In our cases, modeled SEDs with $(\chi^2-\chi_\mathrm{best}^2)<3N_\mathrm{data}$ are treated to be valid. Here, $N_\mathrm{data}$ is the number of data points. The relative probability of each model is estimated to be
    \begin{equation}\label{eq-wei}
      \mathrm{P}(\chi^2)=e^{-(\chi^2-\chi_\mathrm{best}^2)/3N_\mathrm{data}}.
    \end{equation}
    Then it is normalized by dividing $P[(\chi^2-\chi_\mathrm{best}^2)<3N_\mathrm{data}]$. From the well-fitted models for each source returned from the online SED-fitting tool, we calculated $\chi^2$-weighted parameters with assigned weights to be the relative probabilities. The resulting mean values and uncertainties of stellar mass, effective temperature, total luminosity, envelope accretion rate, disk mass, and age of each source are given in Table \ref{tb-sed}. We note that there are sources with thousands of well-fitted models. Such sources always only have data in three adjacent bands. And the fitted parameters are always far from convergent. In the final list given in Table \ref{tb-sed}, 12 sources with more than 3000 well-fitted models ($(\chi^2-\chi_\mathrm{best}^2)<3N$) have been rejected.

    Although the models of \citet{rob06} have been widely used to investigate properties of YSOs, there still are several caveats and limitations that should be borne in mind \citep{rob08}. Generally, the successful fitting of a number of models to an observed SED does not prove that any of these models are actually the correct ones for the object in question, only that they are consistent with the observations \citep{rob08}. And there would be relatively large uncertainties for some parameters, especially the indirectly observable parameters \citep{rob08}. For the stellar mass and age, uncertainties could originate from i) potentially wrong evolutionary tracks of YSOs used to derive some indirect parameters, ii) unresolved multiplicities, and iii) confusion induced from relatively poor resolutions at long wavelengths \citep{rob08}. In this study, the absence of optical photometric data would induce additional uncertainties to some stellar parameters (e.g., age and mass). Similarly, the absence of far-infrared data would lead to high uncertainties to envelope parameters. As mentioned in \citet{rob08}, envelope parameters are also affected by the accuracy of dust models used in \citet{rob06}.

\begin{figure*}
    \centering
    \includegraphics[width=0.95\textwidth]{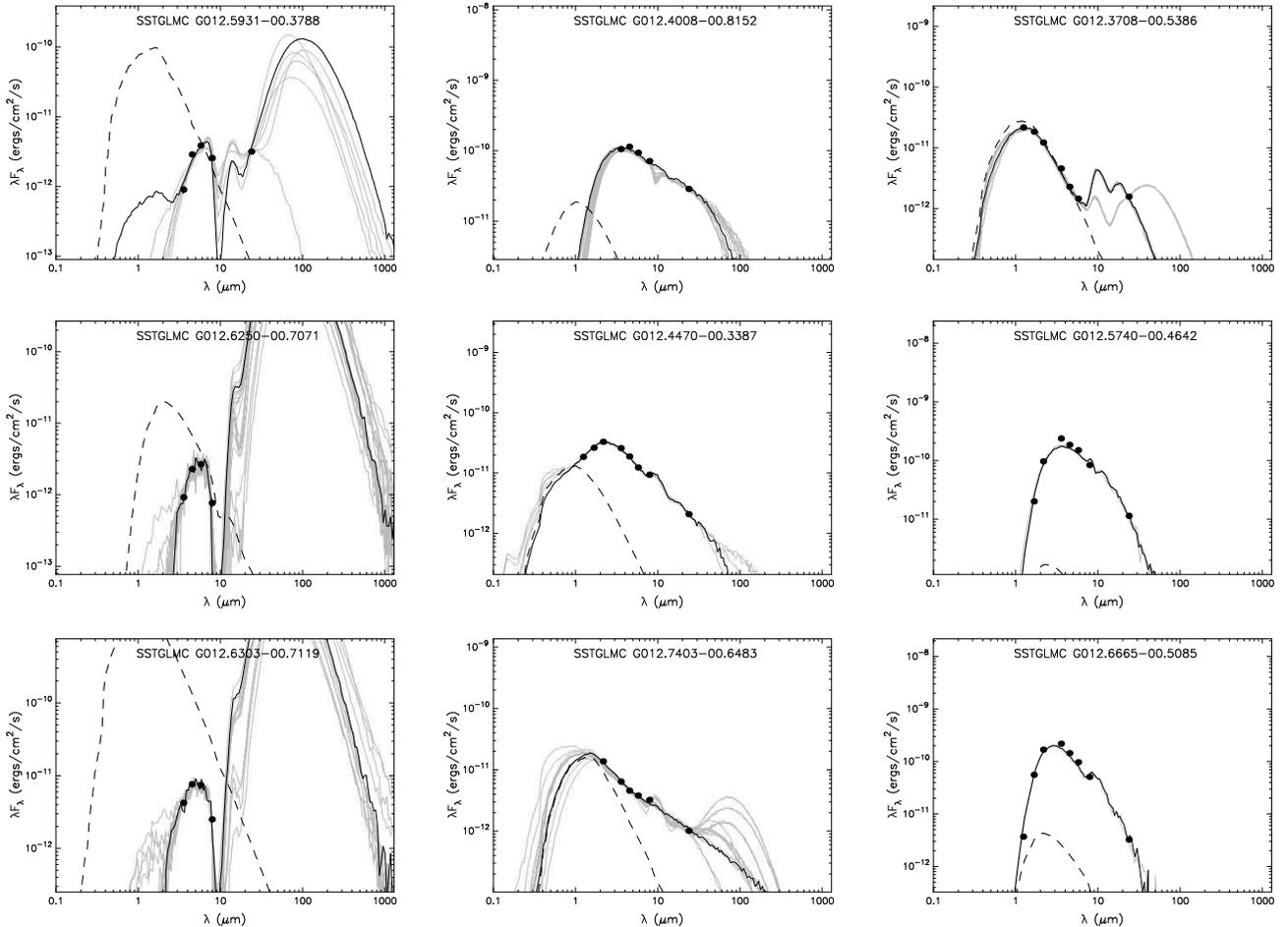}
    \caption{Example SEDs of Stage 0/I (\emph{left}), Stage II (\emph{middle}), and Stage III (\emph{right}) YSOs. The filled circles show the input fluxes. The black lines show the best fit, and the gray lines show subsequent good fits with $(\chi^2-\chi_\mathrm{best}^2)<3N$. The dashed lines show the stellar photosphere corresponding to the central source of the best fitting model, as it would look in the absence of circumstellar dust (but including interstellar extinction) \citep{rob07}.} \label{fig:SEDs}
    \end{figure*}

    For the 87 sources with derived model parameters, we have followed \citet{rob06} to classify them into three categories. In discussing evolutionary stages of their models, \citet{rob06} suggested a ``Stage" classification scheme of YSOs based on their physical properties rather than slopes of the SEDs. With significant infalling envelopes and possible disks, Stage 0/I objects have $\dot{M}_\mathrm{env}/M_\ast>10^{-6}$ yr$^{-1}$. With optically thick disks and possible remains of a tenuous infalling envelope, Stage II objects have $\dot{M}_\mathrm{env}/M_\ast<10^{-6}$ yr$^{-1}$ and $M_\mathrm{disk}/M_\ast>10^{-6}$. With optically thin disks, Stage III sources have $\dot{M}_\mathrm{env}/M_\ast<10^{-6}$ yr$^{-1}$ and $M_\mathrm{disk}/M_\ast<10^{-6}$. The 87 sources are eventually grouped into 25 Stage 0/I, 48 Stage II, and 14 Stage III objects. Example SEDs of several YSOs are presented in Figure \ref{fig:SEDs}.

\section{Discussions}\label{s-discussions}

    Our data in hand have revealed complicated situations and active star-forming activities in N6. In this section, more in-depth analysis of the results will be presented. We will try to constrain the potential exciting stars, address the status of clumps, elucidate the kinematics of the ring structure, and unveil the star-forming scenario.

    \subsection{Exciting Stars}\label{sec:exciting}

    As suggested by \citet{chu06}, most of bubbles have been excited by OB stars. The stars that create bubbles also ionize the ambient neutral material to result in \textsc{H ii} regions which give free-free emission. Radio continuum observations can provide information of such emission and help us constrain the type of stars responsible for the creation of bubbles and the enclosed \textsc{H ii} regions.

    We have measured the total flux at 90 cm above $3\sigma$ ($1\sigma=5$ mJy beam$^{-1}$) inside N6 to be about 8.2 Jy. Given that the bubble is broken, ionized gas may have leaked into the surroundings. Thus, we also measured the 90 cm flux in the near reach of N6 to get a larger value of 16.2 Jy. And we used it as an upper limit, while the value inside N6 as a lower limit. Under an assumption of optically thin in the radio continuum, a relationship between the total flux at a frequency and the ionizing photon rate was proposed by \citet{meg74} as bellow:
    \begin{equation}
    \left(\frac{N_{\rm Ly}}{\rm s^{-1}}\right)\approx4.761\times10^{48}a(\nu,T_\mathrm{e})^{-1}\left(\frac{\nu}{\rm GHz}\right)^{0.1}\left(\frac{T_e}{\rm K}\right)^{-0.45}\left(\frac{S_\nu}{\rm Jy}\right)\left(\frac{D}{\rm kpc}\right)^2.
    \end{equation}
    Here, $a(\nu,T_\mathrm{e})$ is a slowly varying function tabulated by \citet{mez67}; for $T_e=10^4$ K and at radio wavelengths, $a(\nu,T_\mathrm{e})=1$. For the present case, we cannot precisely derive the electron temperature with the data in hand. With an assumed electron temperature of $T_e=10^4$ K, the required ionizing photons per second for the ionization should be in the range of $(6.8-13.3)\times10^{48}$ s$^{-1}$. This indicates that at least one O6.5 star, or a group of smaller ones are needed to produce such a flux \citep{mar05}. But the brightest one should not be earlier than O5.5.

    In the following, we try to constrain possible exciting stars responsible for the creation of N6. The candidates are restricted in the large circle presented in Figure \ref{fig:composite}(b).
    Serving as the source of ionizing photons, the exciting stars would be in main sequence phase. They should have not been identified as YSOs in Section \ref{sec:YSOs}, and have valid detections in all three 2MASS bands. There are 241 point sources fulfilling our requirements. We have followed \citet{par11} to obtain the absolute \emph{J}, \emph{H}, and \emph{Ks} magnitudes. The near infrared extinction law of \citet{rie85} and intrinsic color $[J-H]_0=-0.11$ of O type star in \citet{mar06} were adopted to derive the interstellar extinction of each sample star. With the distance of 3.5 kpc, absolute \emph{J}, \emph{H}, and \emph{Ks} magnitudes of each source were calculated, and then compared with theoretical models in \citet{mar06} to identify potential O type stars which have absolute $Ks$ magnitudes in the range of -5 to -3. We further restrict the candidates using a criterion of $([H-Ks]_0+0.1)<0.15$ based on the theoretical value of $[H-Ks]_0=-0.1$ considering an uncertainty of 0.15 \citep{mar06}. Finally, eight O type stars meeting the above constraints are resulted and tabulated in Table \ref{tb-exciting-star}. These sources have spectral types from O9.5 to O4. However, based on our estimate of ionizing photon flux, the exciting stars should not be brighter than O5.5. Thus, source $\sharp2$ is ruled out.

    We have marked these eight sources in Figure \ref{fig:composite}(b) with small circles. We note that candidates $\sharp$5 and $\sharp$6 are close to the central region of the bubble and could be more likely exciting sources. Additionally, there are some small protuberances on the south-eastern ring-like PAH-dominated structure. Such small structures could be carved by the radiation pressure of the central hot stars. Most of them are pointing to the locations of sources $\sharp$5 and $\sharp$6. This additionally supports that source $\sharp$5 and $\sharp$6 are more likely exciting stars. Further evidence to this conclusion comes from the appearance of the bright-rimed structure associated with clump B, which also points to sources $\sharp$5 and $\sharp$6.

    However, we still cannot rule out the possibility for these exciting star candidates to be giants. Based on the intrinsic colors of giants given in \citet{bes88} and the near infrared extinction laws of \citet{rie85}, we have deduced that the 2MASS colors for a giant fulfill the following relationship of $[J-H]=A[H-Ks]+B$, where $A=\frac{A_J-A_H}{A_H-A_{Ks}}=1.7$, and $B = [J-H]_0-A[H-Ks]_0$ is in the range of 0.26 to 0.43. Among the eight exciting star candidates, six sources have 2MASS colors resembling that of giants. Thus they could also be foreground giants. Nevertheless, there are two possible ionizing stars not contaminated by giants. They are sources $\sharp1$ and $\sharp5$. All of the features mentioned above support that source $\sharp5$ could be the most likely exciting star responsible for the creation of N6.

    There is another bright point source located in the center of N6 as shown in  Figure \ref{fig:composite}. This point source is saturated in all four IRAC bands and not given in the GLIMPSE catalog. It has a 2MASS association with $J$, $H$, and $Ks$ apparent magnitudes of 9.739, 7.124, 5.609. Assuming this source is associated with N6, the absolute magnitude at the $Ks$ band is smaller than -7 mag, even brighter than an O3 type star. Thus this source could be a foreground field star.

    \subsection{Evolutionary Status of Molecular Clumps}

    In Section \ref{sec:clumps}, we have identified ten molecular condensations as clumps and calculated their densities and masses. These clumps have radii ranging from 0.8 to 2.1 pc, H$_2$ column densities ranging from $0.53\times10^{21}$ to $3.86\times10^{22}$ cm$^{-2}$, and masses ranging from 24 to higher than 5,000 $M_\sun$. Such difference in these primary physical parameters between different clumps signifies diverse evolutionary stages.

    To ascertain the dynamical status of these ten clumps, we have followed \citet{mac88} to estimate their virial masses with an assumption of constant density distributions:
    \begin{equation}
    \frac{M_{vir}}{\mathrm{M_\odot}}=210\left(\frac{\Delta v}{\mathrm{km\ s^{-1}}}\right)^2 \left(\frac{R}{\mathrm{pc}}\right).
    \end{equation}
    Here, $\Delta v$ is the width of an observed molecular line and $R$ is the effective radius. For the five clumps with \tco\ detections, line widths of \tco\ $J=1-0$ have been used. For the other five clumps, line widths of \co\ $J=1-0$ have been used. The resylting virial masses are given in Column 15 of Table \ref{tb-cl-der}. Due to the optically thick feature of \co\ $J=1-0$, the derived virial masses for clumps F-J are upper limit values.

    We note that clumps A, C-E, and I have smaller virial masses than the measured gas masses, indicating that they are gravitationally unstable and have potential to collapse to form new stars. Clump B has a virial mass commensurate its LTE mass, indicative of a virial equilibrium status. For the other four clumps, the virial masses are one order of magnitude larger than their non-LTE masses.

    {\bf Clump A} is elongated along NW-SE in \co. It resides on the border of the ring structure. The origin of this clump could be attributed to the fragmentation of the collected shell. Clump A is dense and massive enough to form massive stars. We have detected more than 1,000 $M_\sun$ molecular material in a 2 pc region. It has a LTE mass higher than its virial mass by a factor of three. There is no YSO candidate associated this clump. Near the peak position of \co\ emission, there is an 8.0 $\micron$ point source which spatially coincides with IRAS 18122-1818 (IRAS 7 in this paper). We note that IRAS 18122-1818 has IRAS colors of ${\rm log}(F_{25}/F_{12})=0.49$ and ${\rm log}(F_{60}/F_{12})=1.17$, resembling that of massive star-forming regions \citep{pal91}. Toward IRAS 18122-1818, \citet{deg01} tried to search for 86 GHz SiO maser but reached negative results.

    \begin{figure}
    \centering
    \includegraphics[width=0.45\textwidth]{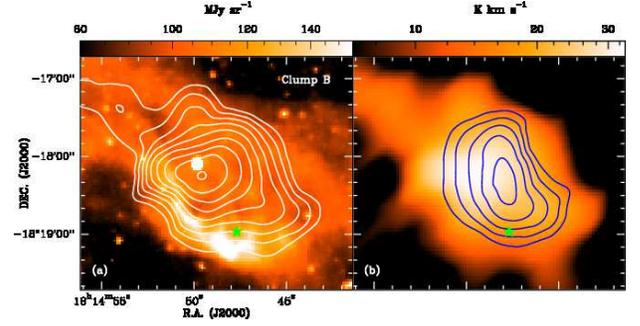}
    \caption{Close-up of clump B. \emph{(a)} Velocity integrated intensity of \co\ $J=1-0$ overlayed on IRAC 8.0 \micron~mosaic. \emph{(b)} Velocity integrated intensity of \tco\ $J=1-0$ overlayed on that of \co\ $J=1-0$ . The filled star represents a Class I type protostar.} \label{fig:cl-B}
    \end{figure}

    {\bf Clump B} is the smallest in size. Near the central portion of this clump, no YSOs have been detected. This indicates that clump B would be at a very early stage of star formation. As shown in Figure \ref{fig:cl-B}, there is an infrared bright rim on the southeast side of this clump. This bright rim is a composition of the ring structure. Intriguingly, the elongation of clump B is along the border of the ring. These features suggest that the expansion of the ring structure is responsible for the formation of clump B. Having equivalent virial and LTE masses, this clump may be on the verge to collapse to form new stars.

    {\bf Clumps C and D} are more elongated than other ones with aspect ratios larger than 2.5. These two clumps are spatially associated with an 8.0 $\micron$ absorbtion structure, which was identified by \citet{per09} as an infrared dark cloud with an identifier of SDC G12.419-0.536. We haven't identified any potential YSO toward the \tco\ peaks of these two clumps. This may suggest that they are in quite early stages prior to the beginning of star formation. However, with LTE and non-LTE masses higher than virial masses, clumps C \& D have potential to collapse to form new stars.

    \begin{figure}
    \centering
    \includegraphics[width=0.45\textwidth]{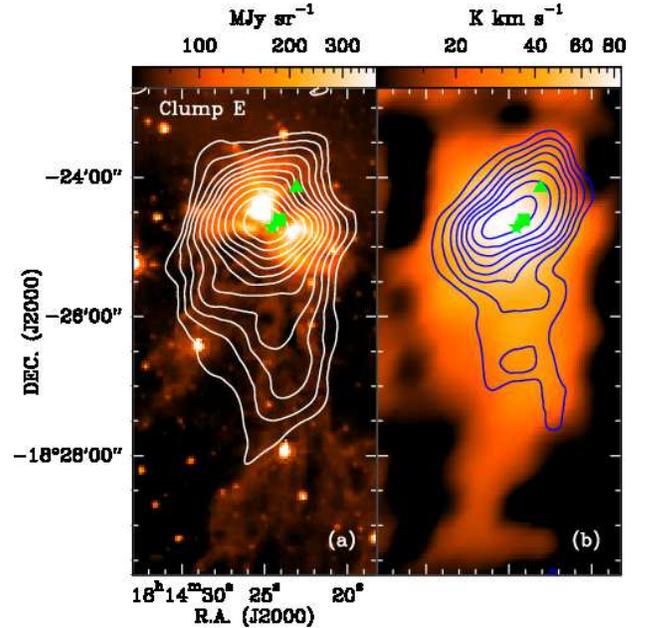}
    \caption{Close-up of clump E. The contours, background are the same as that in Figure \ref{fig:cl-B}. The filed square, triangle, and star represent an IRAS, a Class II, and a Class I sources.}\label{fig:cl-E}
    \end{figure}

    {\bf Clump E} is the densest and most massive one. The column density is higher than $3\times10^{22}$ cm$^{-2}$ and the LTE mass is high up to more than 5,000 $M_\sun$. Such high density and mass assign it to be a massive star-forming clump. As shown in Figure \ref{fig:cl-E}, there is a protostar associated with clump E. It is spatially consistent with the molecular emission peak. This clump shows strong PAH emission features. A symmetric hourglass-shaped bipolar structure have been revealed in the 8.0 $\micron$ band. The center of this bipolar structure is coincident with the peak of molecular emission. This structure has been proposed to be a massive outflow system by \citet{yua12} based on GLIMPSE data. The driving source of this outflow is a Class I YSO, SSTGLMC G012.4013-00.4687. Toward this source, \citet{lim12} have detected an H$_2$ and a 22 GHz water maser emission features supporting outflow activities therein. However, we are unable to resolve this outflow because of the relatively low spatial resolution of our observations. There is an IRAS point source, IRAS 18114-1825 (IRAS 3 in this paper), associated with the bipolar structure and clump E. IRAS 18114-1825 has extraordinarily large infrared luminosity and IRAS colors fulfilling characteristics of a massive star-forming region \citep{pal91}.

    {\bf Clumps F-J} have no reliable detections in \tco. Among these five clumps, only clump I has a non-LTE mass larger than its virial mass. The other four clumps have volume densities lower than $10^3$ cm$^{-3}$. Clumps G, H, and I are distributed along the border of the ring structure. There is IRAS 18123-1829 (IRAS 8) associated with clump H. IRAS 18123-1829 has an infrared luminosity of larger than 3,800 $L_\sun$ and IRAS colors resembling that of massive star-forming regions \citep{pal91}.

    Although the distribution of YSOs suggests most of the ten clumps are not in the place of active star formation, they still could be in different stages. Clump E would be the most evolved one associated with a protostar and showing features of outflow activities. Associated with an IRAS point sources, clumps A  also shows a sign of star formation. Clumps B and I have measured masses larger than their virial masses, and they could be on the verge to collapse to form new stars. Residing in an infrared cloud, clumps C and D could be in an earlier stage. However, they have LTE masses larger than their virial masses and still have potential to give birth to new stars. Among the ten clumps, F, G, H, and J are the most diffuse ones. They are still gravitationally unbound systems. These four clumps need exterior pressures to help them collapse. However, it is also possible for them to dissipate in the long run.
    \subsection{An Expanding Ring}

    As shown in Figure \ref{fig:composite}, there is a ring structure showing bright emission at 8.0 $\micron$. This ring has been reproduced by our PMO observations. In Figure \ref{fig:yso-irac}(a), we present the combined velocity integrated intensity (contours) of the 16.0 and 45.7 km s$^{-1}$ components of \co\ $J=1-0$ overlayed on the {\it Spitzer} 8.0 $\micron$ image. Conspicuously, the ring structure can be well outlined by the distribution of clumps A, B, G, H, and I. As aforementioned in Section \ref{sec:mole}, clumps A and B have line-of-sight velocity of about 16 km s$^{-1}$, while clumps G, H, and I are revealed at about 45.7 km s$^{-1}$. The magnificent association of these clumps with the ring traced by 8.0 $\micron$ suggests that the 16.0 and 45.7 km s$^{-1}$ components are both from a same structure. The velocity difference of these two components could be explained with a hypothesis that this ring structure is expanding with an inclination relative to the plane of sky. Compared with the systemic velocity, the north and south portions of the ring are blue and red shifted, respectively.

    To produce such an expanding ring, the exciting star(s) may be born in a natal molecular cloud with flattened geometry. Such flattened clouds with enclosed ring-like bubbles have been well studied by \citet{bea10}. They suggested that bubbles are formed in parent molecular clouds with a typical thickness of a few parsecs. Additionally, numerical studies have proposed that sheet-like molecular clouds may be natural consequences of ISM evolution in the Galaxy \citep[][ and references therein]{hei05,vaz06}.

    For the case in N6, an O type star (or cluster) formed in an inclined oblate molecular cloud with a limited thickness. The H \textsc{ii} region generated by the central O type star(s) expanded to yield a bubble. When the diameter of the bubble reached the thickness of the natal cloud, the bubble exploded along the flattened axis to result in a ring. Such explosion would lead to the ionization pressure to drop sharply and cause the expansion to be halted. However, as suggested by \citet{bea10}, the ring structure can further expand under the influence of strong winds from the exciting star(s). Simulations of H \textsc{ii} regions carried out by \citet{cap01} and \citet{fre03,fre06} indicates that stellar winds from the exciting star(s) would strengthen with time. If this is true, the ring structure in N6 can hold on its expansion.

    \citet{liu12} presented a similar expanding ring driven by a Wolf-Rayet star. They developed a schematic model to explain the geometry. Based on assumed isotropic stellar winds, the elliptic morphology of the ring in \citet{liu12} was successfully elucidated. Compared with the one in \citet{liu12}, the ring in this paper cannot rely on isotropic stellar winds to maintain the expansion. Because, isotropic stellar winds would lead to a velocity difference along the minor axis of the projected ring. This is conflict with the situation observed in N6, where we have detected prominent velocity difference along the major axis (NW-SE direction). However, this phenomenon can be straightforwardly explained if the stellar winds are mainly constrained along the NW-SE orientation. With an assumed inclination angle of 45$\deg$, we estimate that the central exciting star(s) started blowing strong winds about $4\times10^5$ yr ago.

    The ring is unclosed in the east. There are two possible origins for this opening. A straightforward explanation comes from the NW-SE oriented stellar winds, which have tore the ring to destroy its completeness. Another possibility is that the natal cloud is inhomogeneous with lower density in the east, so that it is easier for the ring to be broken there.

    \subsection{Star Formation in N6}\label{sec:SF}
    \begin{figure}[t]
    \centering
    \includegraphics[width=0.45\textwidth]{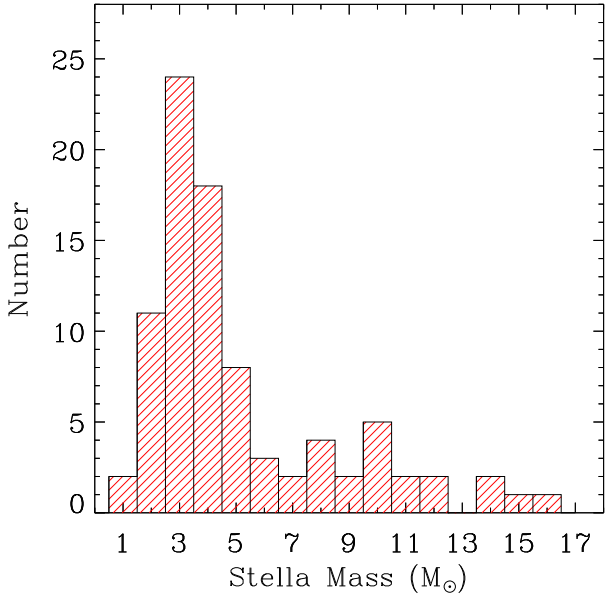}
    \caption{The number of YSOs as a function of stellar mass.}\label{fig:yso-mass}
    \end{figure}

    The 84 YSOs with valid modeled parameters have stellar masses ranging from 1 to 16 $M_\odot$. The number as a function of stellar mass is shown in Figure \ref{fig:yso-mass}. We note that the number of YSOs $M_\ast<3$ $M_\odot$ decreases sharply, indicating that sources with $M_\ast<3$ \msun~are far from complete in our sample. Given that massive stars always disperse their disks quickly, massive YSOs could be incomplete either. Thus, we speculate that sources with masses in the range of 3-8 \msun~could be relatively more complete than both low- and high-mass ones.

    Although most YSO candidates are dispersively distributed, there still reside a small group of YSOs inside the ring structure (see Figure \ref{fig:yso-irac}). Most of them concentrate in the northwest of the cavity surrounded by bright 8.0 \micron~emission. Among the 17 YSOs inside the green ellipse in Figure \ref{fig:yso-irac}(b), 15 have valid SEDs. And nine of them have masses in the range of 3-8 \msun. With an assumed Shalpeter mass function of $dN/d\mathrm{log}(M_\ast)\propto M_\ast^{\Gamma}$ with $\Gamma=-1.35$, there could be 1209 YSOs with masses higher than 0.1 \msun~inside the bubble. Given that mass functions don't necessarily to be universal in distinct regions \citep[][and reference therein]{off13}, the number derived here could be with large uncertainty. In studies of M17 and W51, \citet{pov09} and \citet{kan09} found that active star-forming regions have shallower YSO mass functions with $\Gamma=-1.2$. If this is the case, the number of YSOs with $M_\ast>0.1$ \msun~would be 770. The corresponding surface density of YSOs in a 36 pc$^2$ region (the area of the green ellipse in Figure \ref{fig:yso-irac}(b)) would be in the range of 21 to 33 pc$^{-2}$. This indicates that active star formation have been taking place inside the bubble. The values of number and surface density of YSOs can only be lower limits. We cannot ascertain the YSOs are 100\% complete even for sources with masses in the range of 3-8 \msun. We treat any unresolved multiple sources as single YSOs, which means that only the most luminous component of a multiple system has been sensitively detected, given that $L_\mathrm{tot}\propto M_\ast^{\alpha}$ with $\alpha>2.5$ for pre-main sequence stars \citep{ber96}. Additionally, bright diffuse mid-IR emission near the border of the ring could drastically reduces the sensitivity of the GLIMPSE point-source detections.

    We have inspected the age distribution of the YSOs to seek for footprints of triggered star formation. The ages of the 87 YSOs with valid SED parameters have been coded in colors in Figure \ref{fig:yso-irac}(b). Although the ages of sources inside the large ellipse in Figure \ref{fig:yso-irac}(b) span from $2.0\times10^4$ yr to $6.0\times10^6$ yr,  no evident age gradient has been detected.  Given the complexity of this region, traces of triggered star formation, if any, would have been obliterated in the long run. On the other hand, the YSOs inside the bubble could have formed spontaneously.

    Although it is not easy to convincingly detect triggered star formation in such a complex region, the emersion of the bubble and the expansion of the enclosed \hii~region have influenced the natal cloud and probably star formation therein. As aforementioned and shown in Figure \ref{fig:yso-irac}(a), dense molecular gas has been accumulated along the ring border. Among the 10 molecular clumps detected in this work, five are distributed on the shell. Four of them are elongated along the ring. And the emersion of these clumps does not distort the structure of the shell. This indicates that they may be formed from collected material which moves at the same velocity as the ionization and shock fronts \citep{deh10}. In spite of no active star formation detected in these five clumps, three of them have potential to collapse to serve as cradles for new stars.

     On the basis of a series of smoothed particle hydrodynamic simulations of ionized-induced star formation in bound and unbound systems, \citet{dal12,dal13} argued that triggered stars tend to reside in bubble walls or pillars, while spontaneously formed stars could be found both inside bubble cavities and along bubble walls. In a recent study, \citet{sam14} has observationally revealed the coexistence of triggered and spontaneous star formation in the Galactic \hii~complex Sh2-90. Among the 99 YSOs identified in this work, seven are well associated with the shell of N6 (see Figure \ref{fig:yso-irac}). Their formation could have been triggered. The other YSOs residing inside the cavity and beyond the shell are most likely formed spontaneously. As argued in \citet{dal13}, the association of a YSO with a bubble wall cannot guarantee it a triggered origin, because spontaneously formed YSOs can also be found in bubble walls. Compared to spontaneously formed ones, the triggered stars are always among the youngest objects with strong accretion \citep{dal13}. We have tentatively obtained the ages of six among the seven YSOs associated with the shell structure via fitting their SEDs. As shown in Figure \ref{fig:yso-irac}(b), most of them are relatively evolved with ages larger than 1 Myr save the one associated with Clump B. As a massive protostar ($M_\ast = 9.96\pm3.32$ \msun), this source is still at a very early stage with a high accretion rate ($\dot{M}_\mathrm{env}=5.9\times10^{-3}$ \msun~yr$^{-1}$) and a small age ($\sim2.0\times10^4$ yr). The association with the bubble wall and a young nature make it the most likely triggered object among the identified YSOs.


\section{Conclusions}\label{s-conclusions}

    We have carried out a multi-wavelength study of the infrared dust bubble N6. Observations in \co\ $J=1-0$ and \tco\ $J=1-0$ obtained with the PMO 13.7-m telescope and public survey data in the infrared to radio allow us to achieve an extensive understanding of molecular conditions and star-forming activities therein.

    The bubble N6 shows a complex morphology in the mid-infrared. It is composed of an extended ring in the northeast and a filament in the southwest. There is a bipolar structure in the interspace of the ring and filament, which has been recognized by \citet{yua12} as an bipolar outflow driven by a massive protostar.

    Observations of \co\ and \tco\ have revealed four velocity components centered at 2.8, 16.0, 34.9, and 45.7 km s$^{-1}$, respectively. Comparison between distributions of each component and the infrared emission of N6 suggests that the later three components are correlated with the bubble. The 16.0 and 45.7 km s$^{-1}$ components originate from the the expanding ring, while the 34.9 km s$^{-1}$ component is tracing the natal cloud. The kinetic distance to N6 has been estimated to be about 3.5 kpc using a systemic velocity of 34.9 km s$^{-1}$.

    Ten molecular clumps have been identified. Because of the poor sensitivity of our PMO observations, only five of them have reliable detections in both  \co\ $J=1-0$ and \tco\ $J=1-0$. They have similar LTE and non-LTE masses ranging from 200 to higher than 5,000 $M_\sun$. The virial masses of these five clumps are smaller than their gas masses, indicating that they are gravitationally bound systems and would collapse to form new stars. The other five ones can only be reliably detected in \co\ $J=1-0$. Using \verb"RADEX" program, non-LTE parameters were estimated. Among these five clumps, only clump I have a non-LTE mass larger than its virial mass.

    Among the ten molecular clumps, five ones (A, B, G, H, and I) are located on the border of the ring structure. And four of them show elongations along the shell. This is consistent with a collect and collapse scenario.

    A velocity gradient along the NW-SE orientation for the ring structure has been evidently detected with PMO observations. This indicates that it is still expanding. We suggest that this ring was produced by an H \textsc{ii} region whose expansion has been halted because of the broken of the ring. Stellar winds from the exciting star(s) have been maintaining the expansion of the ring.

    On the basis of infrared color indices, 99  YSO candidates have been identified and classified into Class I, Class II, and ``transition disk" categories. Spectral energy distributions of 87 YSOs were successfully fitted using an online tool. A group of 17 YSOs reside inside the bubble. Based on the number of sources with masses in the range of 3-8 \msun, the lower limits of number and surface density of YSOs are estimated to be 770 and 21 pc$^{-2}$, respectively. This indicates that active star formation has been taking place in the bubble.

    Although no convincing traces of induced star formation have been detected in the bubble, the morphology and kinematics of molecular gas still reveals the influence of the feedback of central hot stars through the expansion of the enclosed \hii~region and intense stellar winds.

\begin{acknowledgements}

    We are grateful to the anonymous referee for the constructive comments that helped us improve this paper. This work is supported by the National Natural Science Foundation of China through grants of 11073027, 11373009 and 11433008, the China Ministry of Science and Technology through grants of 2012CB821800 (a State Key Development Programme for Basic Research) and 2010DFA02710 (by the Department of International Cooperation), Beijing Natural Science Foundation through the grant of 1144015, and the Young Researcher Grant of National Astronomical Observatories, Chinese Academy of Sciences. We give our thanks to the staff at the Qinghai Station of Purple Mountain Observatory for their hospitable assistance during the observations and the Key Laboratory for Radio Astronomy of Chinese Academy of Sciences for partial support in the operation of the telescope.

    This research has made use of the NASA/IPAC Infrared Science Archive, which is operated by the Jet Propulsion Laboratory, California Institute of Technology, under contract with the National Aeronautics and Space Administration. This work is based in part on observations made with the \emph{Spitzer Space Telescope}, which is operated by the Jet Propulsion Laboratory, California Institute of Technology under a contract with NASA. This publication also makes use of data products from the Wide-field Infrared Survey Explorer, which is a joint project of the University of California, Los Angeles, and the Jet Propulsion Laboratory/California Institute of Technology, funded by the NASA. This research made use of APLpy and Astropy for visualization and some analysis.   APLpy is an open-source plotting package for Python hosted at http://aplpy.github.com. And Astropy is a community-developed core Python package for Astronomy \citep{ast13}.

\end{acknowledgements}

    \begin{turnpage}
\begin{deluxetable*}{cccrrrrrrrrrrrrr}
    \tablecaption{Molecular Clumps: Observational Results \label{tb-cl-ob}}
    \tablewidth{0pt}
    \tablehead{ & & & \multicolumn{3}{c}{Deconvolved Size} & &\multicolumn{4}{c}{$^{12}$CO (J=1-0)} &&\multicolumn{4}{c}{$^{13}$CO (J=1-0)} \\
    \cline{4-6}\cline{8-11}\cline{13-16} \\
    ID &
    RA & DEC & \colhead{D$_{major}$}&
    \colhead{D$_{minor}$}&P.A.&&\colhead{$\int T^\ast_{r}dv$} &$T^\ast_{r}$&$v_{lsr}$&FWHM&&$\int T^\ast_{r}dv$ &$T^\ast_{r}$&$v_{lsr}$&FWHM\\
    & (J2000)      &   (J2000)     & (arcsec) &(arcsec)&(degree)&&(K km s$^{-1}$)  &   (K)  &      (km s$^{-1}$)      &   (km s$^{-1}$)     & &  (K km s$^{-1}$)  &   (K)    &      (km s$^{-1}$)    &   (km s$^{-1}$)
    }
    \startdata
    A	&	 18h15m10.49s	& -18d17m31.2s	&	 200.38 & 92.68 & -70.1 && 23.65(2.62)& 12.96 &15.77(0.09) & 1.71(0.18) & & 14.04(1.36)&9.64 & 15.52(0.05) & 1.37(0.16)	  \\

    B	&	 18h14m48.14s	& -18d18m15.6s	&	 111.63 & 64.19 & 30.9 && 26.87(1.51) & 16.37 & 16.27(0.04) &  1.54(0.11)   && 10.38(0.81) & 8.10 & 16.15(0.05) &  1.21(0.10)     \\

    C	&	 18h14m27.51s	& -18d25m09.0s	&	 410.92 & 127.07& 51.4 && 22.37(1.06) & 12.96 &  15.80(0.04) &  1.62(0.08)   && 11.21(0.79) & 5.03 & 15.56(0.07) &  2.09(0.18)        \\

    D	&	 18h14m44.56s	& -18d25m45.6s	&	 185.51 & 69.01 & 71.2 && 18.32(0.94) &13.87&  16.64(0.03) &  1.24(0.08)&&11.22(1.11) & 7.83 &16.52(0.06) &  1.35(0.18)      \\

    E	&	 18h14m24.28s	& -18d24m35.5s	&	 288.32 & 141.89& -10.2	 && 87.10(2.06)  &26.23&  46.04(0.04) &  3.12(0.09)  &&28.74(0.94)   &12.83&46.03(0.03)  & 2.10(0.08)\\

    F    &       18h14m45.54s   &  -18d27m03.2s &   217.16 &  130.99 & -46.4  && 35.79(1.86) & 12.25 & 45.42(0.07) &  2.75(0.16)   && \nodata & \nodata & \nodata & \nodata \\

    G   &        18h14m56.68S   & -18d27m35.4s & 131.35 & 99.39& 27.7 && 35.02(2.89) & 9.28& 45.54(0.13)  & 3.55(0.40) &&   \nodata & \nodata & \nodata & \nodata \\

    H  &         18h15m17.80s   &  -18d29m10.9s  & 193.64 & 127.21 & 49.8 && 33.36(1.56) & 12.75 & 46.80(0.06) &  2.46(0.14) &&   \nodata & \nodata & \nodata & \nodata \\
\enddata
    \end{deluxetable*}
    \end{turnpage}

    \begin{turnpage}
    \begin{deluxetable*}{cccccccccccccccc}
    \tablecaption{Molecular Clumps: Derived Parameters \label{tb-cl-der}}
    \tablewidth{0pt}
    \tablehead{ & & \multicolumn{6}{c}{LTE} & &\multicolumn{6}{c}{non-LTE} & \\
    \cline{3-8}\cline{10-15} \\
    ID & \colhead{R} & $\tau(^{12}\rm{CO})$ & $\tau(^{13}\rm{CO})$ & $T_{\rm{ex}}$ & $N_{\rm{H_2}}$ & $n_{\rm{H_2}}$ & $M_{\rm{LTE}}$& & $T_{\rm{kin}}$ &  $N_{\rm{H_2}}$ & $n_{\rm{H_2}}$ & $T_{\rm{ex}}(^{12}\rm{CO})$ & $T_{\rm{ex}}(^{13}\rm{CO})$ & $M_{\rm{non-LTE}}$ & $M_{\rm{vir}}$\\
    &    (pc)    &   &   &(K)&($10^{21}$ cm$^{-2}$)&($10^{3}$ cm$^{-3}$) &($M_\odot$)  & &  (K)  &   ($10^{21}$ cm$^{-2}$)&($10^{3}$ cm$^{-3}$)  & (K)     & (K) & ($M_\odot$)&($M_\odot$)
    }
    \startdata
    A  &     1.16   &    72.18   &     1.36   &    16.41   &    17.29 (1.67) &     2.42 (0.34) &     1054 (349) & &   15.84 (1.33) &    26.55 (19.22) &     3.72 (2.67) &    15.77 (1.28) &    14.24 (0.72) &     1619 (1258) &      456 (116) \\
B  &     0.72   &    36.19   &     0.68   &    19.85   &    10.69 (0.83) &     2.41 (0.31) &      252 (82) &  &  20.37 (1.34) &     8.54 (1.57) &     1.93 (0.30) &    19.82 (1.14) &    15.70 (0.91) &      201 (68) &      221 (43)      \\
C  &     1.94   &    26.03   &     0.49   &    16.41   &     9.54 (0.67) &     0.80 (0.10) &     1636 (530) & &   16.81 (0.80) &     9.68 (1.51) &     0.81 (0.10) &    16.31 (0.69) &     9.84 (0.59) &     1660 (536) &     1778 (354)   \\
D  &     0.96   &    44.06   &     0.83   &    17.33   &    11.43 (1.13) &     1.93 (0.27) &      480 (159) &  &  17.34 (1.30) &    10.69 (2.26) &     1.80 (0.34) &    17.14 (1.22) &    13.59 (0.87) &      449 (159) &      367 (105)   \\
E  &     1.72   &    35.60   &     0.67   &    29.75   &    38.55 (1.26) &     3.64 (0.38) &     5177 (1646) & &   30.85 (1.19) &    26.75 (3.53) &     2.53 (0.22) &    29.71 (0.99) &    25.49 (1.02) &     3593 (1122) &     1589 (200) \\
    F  &  1.44  &\nodata&\nodata&\nodata&\nodata     &\nodata     &\nodata   && \nodata & 1.75  & 0.21 & \nodata & \nodata & 176  & $\leq$2276\\
    G  &  0.97  &\nodata&\nodata&\nodata&\nodata     &\nodata     &\nodata   && \nodata & 0.85  & 0.15 & \nodata & \nodata & 39   & $\leq$2562\\
    H  &  1.33 &\nodata&\nodata&\nodata&\nodata     &\nodata     &\nodata   && \nodata & 1.89  & 0.28 & \nodata & \nodata & 164  & $\leq$1692\\
\enddata
    \end{deluxetable*}
    \end{turnpage}

    \begin{deluxetable*}{ccrrrccccr}
    \tablecaption{IRAS Sources \label{tb-iras}}
    \tablewidth{0pt}
    \tablehead{
    &IRAS Name & \colhead{$F_{12}$} & $F_{25}$ & $F_{60}$ & $F_{100}$ & f$_\mathrm{qual}\tablenotemark{a} $& log($\frac{F_{25}}{F_{12}}$) &  log($\frac{F_{60}}{F_{12}}$) & \colhead{$L_{IR}$}  \\
      &         &    (Jy)  &    (Jy)  &    (Jy)  &    (Jy)  &           &                 &(L$_\odot$)
    }
    \startdata
    IRAS1  & IRAS 18114-1817	&	1.6	&	6.2	 &	27.4	 &	296.2	& 3111 &	 0.59	 &	 1.24	  	 &	 2805	 \\
    IRAS2  & IRAS 18114-1828	&	2.5	&	3.0	 &	24.0	 &	659.8	& 1311 &	 0.07	 &	 0.98	  	 &	 5167	 \\
    IRAS3  & IRAS 18114-1825	&	7.2	&  22.1	 &	339.0	 &	659.8	& 2323 &	 0.49	 &	 1.68	  	 &	 11586	 \\
    IRAS4  & IRAS 18118-1828	&	2.0	&	2.2	 &	40.7	 &	184.6	& 3121 &	 0.03	 &	 1.31	  	 &	 2182	 \\
    IRAS5  & IRAS 18120-1834	&	2.5	&	2.6	 &	52.4	 &	645.4	& 3131 &	 0.02	 &	 1.32	  	 &	 5543	 \\
    IRAS6  & IRAS 18121-1825	&  15.1	&  12.9	 &	31.4	 &	623.9	& 3311 &	 -0.07	 &	 0.32	  	 &	 6446	 \\
    IRAS7  & IRAS 18122-1818	&	3.9	&  12.1	 &	57.9	 &	623.8	& 2311 &	 0.49	 &	 1.17	  	 &	 5926	 \\
    IRAS8  & IRAS 18123-1829	&	4.2	&	5.5	 &	75.3	 &	302.1	& 1131 &	 0.12	 &	 1.26	  	 &	 3861	 \\
    IRAS9  & IRAS 18126-1820	&	2.0	&	4.2	 &	24.4	 &	377.8	& 1311 &	 0.33	 &	 1.10	  	 &	 3266
    \enddata
    \tablenotetext{a}{Flux density quality: 1 for an upper limit, 2 for moderate quality and 3 for high quality. The four figures for each source refer to the flux density quality of \emph{IRAS} bands at 12, 25, 60 and 100 \micron, respectively.}
    \end{deluxetable*}

    \begin{deluxetable*}{lrr}
    \tablecaption{Source Counts in the YSO Search \label{tb-count}}
    \tablewidth{0pt}
    \tablehead{
      \colhead{Sources} &   \colhead{N6 Target Field}    &   \colhead{Control Field}
    }
    \startdata
    In GLIMPSE Catalog              & 52737  & 42894 \\
    Valid in all Four IRAC Bands\tablenotemark{a}  & 7943   & 9078   \\
    With 24 \micron~data            & 491    & 534   \\
    Rejected Contaminants           & 2474  & 2332   \\
    Class I Type                    & 27     & 11   \\
    Class II Type                   & 48     & 19   \\
    Transition Disks                & 24     &  9   \\
    Total YSO Candidates            & 99     & 39
    \enddata
    \tablenotetext{a}{Sources with photometric uncertainties no larger than 0.2 mag in all four IRAC bands.}
    \end{deluxetable*}

\clearpage
\begin{landscape}
   \LongTables

    \begin{deluxetable}{rrrrrrrrrrrrc}
    \tablecaption{Magnitudes of YSO Candidates \label{tb-photo}}
    \tablewidth{0pt}
    \tablehead{
    \colhead{No.} & \colhead{GLIMPSE Designation} & \colhead{R.A.} & \colhead{Dec} & \multicolumn{8}{c}{Magnitudes} & \colhead{Class} \\
    \cline{5-12} \\
     & & \colhead{(deg)} & \colhead{(deg)} & \colhead{J}& \colhead{H}& \colhead{Ks}& \colhead{[3.6]}& \colhead{[4.5]}& \colhead{[5.8]}& \colhead{[8.0]}& \colhead{[24]} &
    }
    \startdata
    1  & SSTGLMC G012.1558-00.4524 &  273.4635 &  -18.6197    &     \multicolumn{1}{c}{\nodata}  &    \multicolumn{1}{c}{\nodata} &    \multicolumn{1}{c}{\nodata}  & 12.19(0.09) &  11.79(0.10)   &    \multicolumn{1}{c}{\nodata}  &    \multicolumn{1}{c}{\nodata}  &   5.90(0.02)   &   TD                        \\
    2  & SSTGLMC G012.1903-00.5208 &  273.5443 &  -18.6221    &     \multicolumn{1}{c}{\nodata}  &    \multicolumn{1}{c}{\nodata} & 11.09(0.06)  &  7.98(0.04) &   6.68(0.04)   &  5.67(0.03)  &  4.94(0.02)  &   3.85(0.01)   &    I                  \\
    3  & SSTGLMC G012.2113-00.4422 &  273.4821 &  -18.5661    &  15.11(0.07)  & 13.77(0.08) & 12.63(0.05)  & 11.12(0.04) &  10.71(0.06)   & 10.35(0.05)  &  9.67(0.05)  &   6.69(0.01)   &   II              \\
    4  & SSTGLMC G012.2166-00.6754 &  273.7009 &  -18.6728    &  13.93(0.04)  & 12.80(0.03) & 11.91(0.03)  & 10.52(0.04) &  10.00(0.05)   &  9.85(0.05)  &  8.70(0.03)  &   6.38(0.01)   &   II              \\
    5  & SSTGLMC G012.2518-00.4213 &  273.4831 &  -18.5205    &     \multicolumn{1}{c}{\nodata}  &    \multicolumn{1}{c}{\nodata} &    \multicolumn{1}{c}{\nodata}  & 13.63(0.14) &  12.98(0.17)   &    \multicolumn{1}{c}{\nodata}  &    \multicolumn{1}{c}{\nodata}  &   6.90(0.06)   &   TD                        \\
    6  & SSTGLMC G012.2552-00.4189 &  273.4826 &  -18.5163    &     \multicolumn{1}{c}{\nodata}  &    \multicolumn{1}{c}{\nodata} &    \multicolumn{1}{c}{\nodata}  & 12.85(0.11) &     \multicolumn{1}{c}{\nodata}   & 10.27(0.07)  &  9.86(0.06)  &   5.37(0.01)   &   TD                      \\
    7  & SSTGLMC G012.2580-00.4010 &  273.4674 &  -18.5053    &     \multicolumn{1}{c}{\nodata}  &    \multicolumn{1}{c}{\nodata} & 11.35(0.03)  &  8.39(0.04) &   6.90(0.04)   &  5.67(0.03)  &  4.96(0.03)  &   3.19(0.01)   &    I                  \\
\enddata
    \end{deluxetable}
\clearpage
\end{landscape}
   \LongTables
    \begin{deluxetable}{rrrrrcccc}
    \tabletypesize{\scriptsize} \tablecolumns{9} \setlength{\tabcolsep}{0.001in}
    \tablecaption{Model Parameters of YSO Candidates \label{tb-sed}}
    \tablewidth{\textwidth}
    \tablehead{
    \colhead{No.} & \colhead{GLIMPSE Designation} & \colhead{$M_\ast$} & \colhead{log($T_\ast$)} &  \colhead{log($L_\ast$)} & \colhead{log($\dot{M}_\mathrm{env}$)} & \colhead{log($M_\mathrm{disk}$)} & \colhead{log($Age$)} & \colhead{Stage} \\
     & & \colhead{($M_\sun$)} & \colhead{(log(K))} & \colhead{(log($L_\sun$))}& \colhead{(log($M_\sun\ \mathrm{yr}^{-1}$))}& \colhead{(log($M_\sun$)}& \colhead{(log(yr))}&
    }
    \startdata
  1 &  SSTGLMC G012.1558-00.4524  &    2.34(1.06)    &    3.71(0.15)    &     1.37(0.36)    &    -5.77(1.34)    &   -2.99(1.54)    &    5.53(0.49)    &     II  \\
  2 &  SSTGLMC G012.1903-00.5208  &   11.47(0.01)    &    4.44(0.01)    &     3.98(0.01)    &    $\leq -9.00$    &   -5.63(0.01)    &    6.26(0.01)    &    III  \\
  3 &  SSTGLMC G012.2113-00.4422  &    2.87(0.05)    &    4.06(0.01)    &     1.79(0.03)    &    $\leq -9.00$    &   -2.63(0.42)    &    6.78(0.16)    &     II  \\
  4 &  SSTGLMC G012.2166-00.6754  &    3.42(0.01)    &    4.11(0.01)    &     2.07(0.01)    &    $\leq -9.00$    &   -4.65(0.01)    &    6.58(0.01)    &     II  \\
  6 &  SSTGLMC G012.2552-00.4189  &    3.04(0.59)    &    3.65(0.01)    &     1.84(0.05)    &    -3.94(0.08)    &   -1.05(0.21)    &    4.90(0.16)    &    0/I  \\
  7 &  SSTGLMC G012.2580-00.4010  &    9.28(0.01)    &    4.39(0.01)    &     3.69(0.01)    &    $\leq -9.00$    &   -5.11(0.01)    &    6.40(0.01)    &    III  \\
  9 &  SSTGLMC G012.2765-00.4952  &    9.75(0.01)    &    4.40(0.01)    &     3.76(0.01)    &    $\leq -9.00$    &   -5.72(0.01)    &    6.25(0.01)    &    III  \\
 10 &  SSTGLMC G012.2830-00.4240  &    4.55(0.83)    &    3.97(0.22)    &     2.34(0.14)    &    -7.17(2.39)    &   -2.09(0.82)    &    5.99(0.64)    &     II  \\
\enddata
    \end{deluxetable}

\begin{deluxetable*}{cccrrrcccccc}
\tabletypesize{\scriptsize}\setlength{\tabcolsep}{0.05in}
\tablecaption{Exciting star candidates\label{tb-exciting-star}}
\tablewidth{0pt}
\tablehead{\colhead{No.} & \colhead{GLIMPSE name} & \colhead{$m_J$}  &  \colhead{$m_H$}  &  \colhead{$m_{Ks}$}&
   \colhead{$A_V$} & \colhead{$M_J$}  &  \colhead{$M_H$}  &  \colhead{$M_{Ks}$} &
   \colhead{$[J-H]_0$}  &  \colhead{$[H-Ks]_0$} & \colhead{Spectral Type}
}
\startdata
       1 & SSTGLMC G012.4601-00.6376  &    10.62 &     9.98 &     9.75 &     7.01 &    -4.08 &    -3.97 &    -3.76 &    -0.11 &    -0.21 &  O7.5-O7	 \\
 2 & SSTGLMC G012.4669-00.5793  &    10.82 &     9.77 &     9.31 &    10.84 &    -4.96 &    -4.85 &    -4.62 &    -0.11 &    -0.22 &  O5-O4   \\
 3 & SSTGLMC G012.4684-00.5713  &    11.54 &    10.73 &    10.41 &     8.60 &    -3.61 &    -3.50 &    -3.27 &    -0.11 &    -0.22 &  O9-O8.5  \\
 4 & SSTGLMC G012.4743-00.5650  &    11.47 &    10.70 &    10.43 &     8.22 &    -3.57 &    -3.46 &    -3.21 &    -0.11 &    -0.25 &  O9.5-O8.5   \\
 5 & SSTGLMC G012.5039-00.5783  &    11.05 &    10.26 &     9.86 &     8.41 &    -4.04 &    -3.93 &    -3.80 &    -0.11 &    -0.13 &  O7.5-O7   \\
 6 & SSTGLMC G012.5133-00.5968  &    15.18 &    12.64 &    11.31 &    24.77 &    -4.52 &    -4.41 &    -4.18 &    -0.11 &    -0.23 &  O6-O5.5  \\
 7 & SSTGLMC G012.5179-00.5474  &    11.45 &    10.73 &    10.47 &     7.76 &    -3.46 &    -3.35 &    -3.12 &    -0.11 &    -0.23 &  O9.5-O9  \\
 8 & SSTGLMC G012.5380-00.6083  &    11.58 &    10.73 &    10.41 &     8.97 &    -3.67 &    -3.56 &    -3.32 &    -0.11 &    -0.25 &  O9-O8.5
\enddata
\end{deluxetable*}

\end{document}